\documentclass[preprint,amsmath,amssymb,superscriptaddress,nofootinbib]{revtex4}

\usepackage{bm}

\usepackage{booktabs}

 \usepackage[pdftex]{graphicx,color}
 \usepackage{epstopdf}
 \usepackage[bookmarks=true,bookmarksnumbered=true,bookmarkstype=toc]{hyperref} 
 \usepackage{units}

\newcommand{\lsim}{\raise0.3ex\hbox{$\;<$\kern-0.75em\raise-1.1ex\hbox{$\sim\;$}}}
\newcommand{\gsim}{\raise0.3ex\hbox{$\;>$\kern-0.75em\raise-1.1ex\hbox{$\sim\;$}}}
\usepackage{mathtools}

\preprint{UME-PP-007}

\usepackage[utf8]{inputenc}

\usepackage{color}

\begin{document}

\title{On the possibility of  search for $L_\mu - L_\tau$ gauge boson at Belle-II and neutrino beam experiments}

\author{Yuya Kaneta}
\email{kaneta@muse.sc.niigata-u.ac.jp}
\affiliation{Graduate School of Science and Technology, Niigata University, Niigata 950-2181, Japan}

\author{Takashi Shimomura}
\email{shimomura@cc.miyazaki-u.ac.jp}
\affiliation{Faculty of Education, Miyazaki University, Miyazaki, 889-2192, Japan}

\begin{abstract}
We study the possibilities on the search of the light and weakly interacting gauge boson in the gauged $L_\mu - L_\tau$ model. 
Introducing the kinetic mixing at the tree-level, the allowed parameter regions for the gauge coupling and kinetic mixing parameter 
are presented. Then, we analyze one photon plus missing event within the allowed region and show that search for the light 
gauge boson will be possible at Belle-II experiment. We also analyze neutrino trident production process in neutrino beam experiments.
\end{abstract}

\date{\today}

\maketitle

\section{Introduction}
The anomalous magnetic moment of the muon, $(g-2)_\mu$, is one of the most precisely measured and calculated quantities 
in particle physics. Therefore it can provide a sensitive search for new physics beyond the Standard Model (SM).
Over recent decades,  there has remained a discrepancy between experimental values 
\cite{Bennett:2006fi, Olive:2016xmw} and 
the SM predictions
\cite{Davier:2010nc,Jegerlehner:2011ti,Hagiwara:2011af,Aoyama:2012wk}, 
\begin{align}
\Delta a_\mu \equiv a_\mu^{\mathrm{exp}} - a_\mu^{\mathrm{theo}}= (28.8 \pm 8.0) \times 10^{-10}, \label{eq:mu-g-2}
\end{align} 
which corresponds to $3.6\sigma$ deviation from the SM prediction. 
The discrepancy can be verified in forthcoming   
experiments that will reduce the uncertainties by about a factor of four \cite{Aoki-j-park, Grange:2015fou}. 
If the discrepancy is confirmed 
by the experiments, it will be a clear evidence of new physics beyond the SM. 

On theoretical side, many extensions of the SM have been proposed to explain this discrepancy 
so far (see \cite{Jegerlehner:2009ry, Lindner:2016bgg} for review and \cite{Fayet:2016nyc} for recent works). 
Among them, new $U(1)$ gauge symmetries are of particular interest since these are one of the minimal extensions of the SM. 
To resolve the discrepancy of $(g-2)_\mu$ in this class of models, the simplest possibility is that muons are charged 
under the new symmetry while other SM particles are neutral. 
Then, the muon receives a contribution from the new gauge boson to its anomalous magnetic moment. 
For the $U(1)$ symmetry to be anomaly-free, the condition, $3B = L_e + L_\mu + L_\tau$, must be satisfied where   
$B$ is the baryon number and $L_{e,\mu,\tau}$ are the flavour numbers, respectively. 

Among anomaly free $U(1)$ symmetries, the $L_\mu - L_\tau$ symmetry is 
particularly interesting \cite{Foot:1990mn, Foot:1994vd, He:1991qd}. The models with the $L_\mu - L_\tau$ symmetry 
can provide the large atmospheric mixing as the leading approximation, with some extensions such as adding 
right-handed neutrinos and new scalar particles to obtain the correct 
reactor angle of the lepton mixing \cite{Choubey:2004hn, Ota:2006xr, Heeck:2010pg, Heeck:2011wj, Heeck:2014qea, Biswas:2016yjr, Biswas:2016yan}, and these also can explain the gap in the cosmic neutrino spectrum observed by 
IceCube \cite{Aartsen:2014gkd,  Araki:2014ona, Kamada:2015era, Araki:2015mya}. 
Furthermore, when the interactions between quarks and the gauge boson associated with the symmetry are 
introduced, the models can explain the anomalies reported by 
LHCb \cite{Altmannshofer:2014cfa, Altmannshofer:2015mqa, Altmannshofer:2016jzy}. 
Recent studies in the model can be found for neutrino trident production processes \cite{Altmannshofer:2014pba, Magill:2016hgc}, 
rare Kaon decays \cite{Ibe:2016dir}, lepton flavour violations \cite{Altmannshofer:2016brv} and 
related phenomenologies \cite{Harigaya:2013twa, Patra:2016shz, Biswas:2016yan, Heeck:2016xwg, Biswas:2016yjr}. 
For direct and indirect searches of such the gauge boson,  new experiments are under preparation \cite{Gninenko:2014pea, Acciarri:2015uup, Anelli:2015pba}.
In the previous studies, the result of \cite{Altmannshofer:2014pba} showed the gauge boson mass and the coupling constant must 
be lighter than $400$ MeV and smaller than $10^{-3}$ without kinetic mixing model. 
Such the light and weakly interacting gauge boson will be difficult to search in high energy experiments because its 
production cross sections and decay branching ratios are very suppressed. 
Therefore, high-luminosity or high-flux experiments like the Belle-II and neutrino oscillation experiments are 
suitable for the search of such the gauge boson. 

The search of light and weakly interacting gauge bosons at the Belle-II experiment has been studied 
in the context of the dark photon scenario \cite{Essig:2009nc, Essig:2013vha}, where the SM fermions interact with the dark 
photon only through the kinetic mixing with the photon. On the other hand, in the studies on $L_\mu - L_\tau$ models 
mentioned above,  the kinetic mixing at tree-level is usually set to be zero by hand. 
Such the tree-level kinetic mixing, however, is allowed by the symmetries and therefore should be considered simultaneously. 
In this paper, we consider a model with the gauged $U(1)_{L_\mu - L_\tau}$ symmetry in the presence of the kinetic 
mixing, and explore the allowed parameter space for the light and weakly interacting gauge boson. 
Then, we study the possibilities of search for such the gauge boson in one-photon plus missing event at the Belle-II experiment, 
and in the neutrino trident production process at neutrino beam experiments.

This paper is organized as follows. 
In section \ref{sec:model}, we introduce the model with the gauged $U(1)_{L_\mu- L_\tau}$ symmetry, and show the relevant interactions 
and decay widths of the $L_\mu- L_\tau$ gauge boson.
In section \ref{sec:constraints}, the experimental constraints and requirements to restrict the model parameters are 
explained. Then, in section \ref{sec:allowed-region}, we show 
the allowed parameter regions of the model. In section \ref{sec:results}, the possibilities of search for the gauge boson 
at Belle-II and neutrino beam experiments are discussed. Section \ref{sec:summary} is devoted to summary and discussions.

\section{Gauged $L_\mu - L_\tau$ model} \label{sec:model}

We start our discussion with introducing our model. 
The SM is extended by adding the gauged $U(1)_{L_\mu - L_\tau}$ symmetry under which muon and 
tau flavour leptons are charged. 
The charge assignment of the symmetry is summarized in Table \ref{charge}. Here $l_{\mu}$ and $l_{\tau}$ 
represent $SU(2)$ doublets, and $\mu_R$ and $\tau_R$ represent $SU(2)$ singlets of muon and tau flavours, respectively.
\begin{table}[t]
  \begin{center}
    \begin{tabular}{|c|c|c|c|c|}\hline
      & $~~l_\mu = (\nu_\mu , \mu_L)^T~~$ & $~~l_\tau = (\nu_\tau , \tau_L)^T~~$ & $~~~~~~~\mu_R~~~~~~~$ & $~~~~~~~\tau_R~~~~~~~$ \\ \hline
      $~~~U(1)_{L_\mu -L_\tau}~~~$ & $1$ & $-1$ & $1$ & $-1$ \\ \hline
    \end{tabular}
  \end{center}
  \caption{The charge assignment of gauged $U(1)_{L_\mu - L_\tau}$ model. 
  Here, $l_{\mu}$ and $l_{\tau}$ represent $SU(2)$ doublets, and $\mu_R$ and $\tau_R$ represent $SU(2)$ 
  singlets of muon and tau flavours, respectively. All other SM fermions and the Higgs are singlet under this symmetry.
}
  \label{charge}
\end{table}
Then, the Lagrangian of the model takes the form of  
\begin{align}
\mathcal{L} &= \mathcal{L}_{\mathrm{SM}} - V_{L_\mu - L_\tau}
  - \frac{1}{4} Z'_{\mu\nu} Z'^{\mu\nu} 
   + \frac{\epsilon}{2} B_{\mu \nu} Z'^{\mu\nu}
   + g' Z'_\mu J_{Z'}^\mu,\\
J_{Z'}^\mu &=
 \overline{l_\mu} \gamma^\mu l_\mu 
 +\overline{\mu_R} \gamma^\mu \mu_R
 -\overline{l_\tau} \gamma^\mu l_\tau
 -\overline{\tau_R} \gamma^\mu \tau_R, \label{eq:current}
\end{align}
where $\mathcal{L}_{\mathrm{SM}}$ and $V_{L_\mu - L_\tau}$ stand for the SM Lagrangian and the scalar 
potential responsible for the $L_\mu - L_\tau$ symmetry breaking, and  $Z'$ and $B$ represent the gauge fields 
of the $U(1)_{L_\mu - L_\tau}$ and the hypercharge $U(1)_Y$, respectively. The same symbols 
for their field strengths are used. The gauge coupling constant and the kinetic mixing parameter 
are denoted as $g'$ and $\epsilon$, and the $U(1)_{L_\mu - L_\tau}$ current, $J_{Z'}^\mu$, is given by 
Eq.\eqref{eq:current}. 
In this work, we concentrate our discussion on the gauge sector and hence do not specify the 
potential $V_{L_\mu - L_\tau}$. 
We assume that the $L_\mu - L_\tau$ symmetry is spontaneously broken without conflicting experimental 
constraints such as the SM Higgs couplings to the SM gauge bosons\footnote{This assumption can be 
realized when we introduce a scalar $S$ which is singlet under the SM gauge symmetries. 
Such the scalar has a quartic interaction with the SM Higgs, $|S|^2 |H|^2$.
However, their mixing can be very small by taking the quartic coupling enough small.}. Therefore,  
we treat the mass of $Z'$, $m_{Z'}$, as a free parameter. 
 
After the electroweak and $L_\mu - L_\tau$ symmetries are broken down, the gauge bosons acquire the masses and 
their neutral components mix each other due to the kinetic mixing. 
Then the interaction Lagrangian of leptons with the gauge bosons in mass-basis is obtained by diagonalizing 
their kinetic terms as well as mass terms. 
Assuming $m_{Z'}$ is much lighter than the $Z$ boson mass, the interaction Lagrangian is 
given by
\begin{align}
\mathcal{L}_{\mathrm{int}} &=
e A_\mu J_{\rm EM}^\mu 
+g_2 Z_\mu J_{\rm NC}^\mu 
+ Z'_\mu \left(
  e \epsilon\cos\theta_W J_{\rm EM}^\mu
  +g' J_{Z'}^\mu
  \right)
  + \mathcal{O}(\epsilon^2), \label{eq:int-lag-mass}
\end{align}
where $J_{\rm EM}^\mu$ and $J_{\rm NC}^\mu$ are the electromagnetic and weak neutral currents of the SM, 
respectively, and $e$ and $\theta_W$ are the electric charge and the Weinberg angle. 
In Eq.~\eqref{eq:int-lag-mass}, we have neglected the terms of order $\epsilon^2$ and higher order ones. Such the terms 
include the interaction of electron neutrinos with $Z'$.
As we will show in the following sections, the kinetic mixing parameter and the gauge coupling constant of our interest 
are smaller than $10^{-3}$. Therefore, these terms can be safely ignored in our discussion. 
The kinetic mixing also can be generated via muon and tau loops, which is two orders of magnitude suppressed than $g'$. 
We also ignore such the kinetic mixing for simplicity. 

The decay widths of $Z'$ are given by,
\begin{subequations}
\begin{align}
\Gamma(Z' \rightarrow \nu \bar{\nu}) &= \frac{g'^2}{24 \pi} m_{Z'}, \\
\Gamma(Z' \rightarrow e^+ e^-) &= \frac{(\epsilon e \cos\theta_W)^2}{12 \pi} m_{Z'} 
   \sqrt{1 - \frac{4 m_e^2}{m_{Z'}^2}} \left( 1 + \frac{2 m_e^2}{m_{Z'}^2} \right), \\
\Gamma(Z' \rightarrow l^+ l^-) &= \frac{(g' \mp \epsilon e \cos\theta_W)^2}{12 \pi} m_{Z'} 
   \sqrt{1 - \frac{4 m_l^2}{m_{Z'}^2}} \left( 1 + \frac{2 m_l^2}{m_{Z'}^2} \right), \label{eq:width-l} \\
\Gamma(Z' \rightarrow \mathrm{hadrons}) &= \frac{(\epsilon  e \cos \theta_W)^2}{12 \pi} m_{Z'} 
   \sqrt{1 - \frac{4 m_\mu^2}{m_{Z'}^2}} \left( 1 + \frac{2 m_\mu^2}{m_{Z'}^2} \right) R(s = m_{Z'}^2),
   \label{eq:width-had}
\end{align} 
\label{eq:decay-width}
\end{subequations}
where $l = \mu,\tau$ and the sign in Eq.~\eqref{eq:width-l} is $-$ for $\mu$ and $+$ for $\tau$, respectively.
In Eq.~\eqref{eq:width-had},  $R(s)$ is the $R$-ratio defined by 
$\sigma_{e^+ e^- \rightarrow \mathrm{hadrons}}/\sigma_{e^+ e^- \rightarrow \mu^+ \mu^-}$ and can be found 
in \cite{Olive:2016xmw}. For $\sqrt{s} \lsim 0.36$ GeV, we use the cross section for $e^+ + e^- \rightarrow \pi^+ + \pi^-$ 
\cite{Ezhela:2003pp, Davier:2002dy}.
The branching ratio of $Z' \rightarrow \nu + \bar{\nu}$ obtained from Eqs.~\eqref{eq:decay-width} is used in the following 
analyses.

\section{Experimental Constraints} \label{sec:constraints}
In this section, we explain experimental bounds and requirements to constrain the parameters of the model, $g'$, $\epsilon$ 
and $m_{Z'}$.

\subsection{Muon anomalous magnetic moment}
As mentioned in the introduction, muons receive contributions from $Z'$ to its anomalous magnetic 
moment. At the one-loop level, the contribution is given by
\begin{align}
\Delta a_\mu^{Z'} = \frac{(g'-\epsilon e \cos\theta_W)^2}{8 \pi^2} \int^1_0 dx \frac{2 m_\mu^2 x^2 (1-x)}{x^2 m_\mu^2 + (1-x)m_{Z'}^2},
\label{eq:delta-mu-g-2}
\end{align}
where $m_\mu$ is the mass of a muon. We require the contribution Eq.\eqref{eq:delta-mu-g-2}
to be within $2\sigma$($3\sigma$), that leads to
\begin{align}
12.8~(4.8) \lsim \Delta a_\mu^{Z'} \times 10^{10} \lsim 44.8~(52.8).
\end{align}

\subsection{Neutrino trident production process}
The neutrino trident production process is the scattering of a muon neutrino off the Coulomb field of a nucleus ($N$),  
producing two muons in the final state, $\nu_\mu + N \rightarrow \nu_\mu + \mu^+ + \mu^- + N$.
This process can occur both in the SM and in the
$L_\mu - L_\tau$ model. The process offers a sensitive search for the light $Z'$ boson as shown 
in \cite{Altmannshofer:2014pba, Magill:2016hgc} since the SM contributions are much suppressed due to the weak 
interaction. 
The experimental search results 
have been reported by CCFR \cite{Mishra:1991bv} and CHARM-II \cite{Geiregat:1990gz} collaborations, and the 
most stringent bound was set by the CCFR experiment, 
\begin{align}
R_{\mathrm{CCFR}} \equiv \frac{\sigma_{\mathrm{CCFR}}}{\sigma_{\mathrm{SM}}} &= 0.82 \pm 0.28. \label{eq:const-ccfr}
\end{align}
In \cite{Altmannshofer:2014pba}, it was shown that the favored parameter region of $(g-2)_\mu$ is excluded for $m_{Z'} \gsim 400$ MeV 
without the kinetic mixing in the $L_\mu - L_\tau$ model. In Sec.~\ref{sec:allowed-region}, 
we calculated the trident production cross section under the equivalent 
photon approximation \cite{vonWeizsacker:1934nji, Williams:1934ad} 
using CalcHEP \cite{Belyaev:2012qa} for photon-neutrino scattering cross section. 
We found that our cross sections and results are in good agreement with \cite{Brown:1973ih} and \cite{Altmannshofer:2014pba}. 
We require the $Z'$ contribution should be less than $95$\% C.L. of Eq.~\eqref{eq:const-ccfr}.

\subsection{Neutrino-electron scatterings}
Neutrino-electron scattering tightly constrains $g'$ and $\epsilon$ for a dark photon and a light $Z'$ 
boson \cite{Harnik:2012ni, Bilmis:2015lja}. 
To our model, the constraints from reactor neutrino experiments, e.g. the TEXONO 
experiment \cite{Deniz:2009mu, Li:2002pn, Wong:2006nx, Chen:2014dsa},  
are irrelevant because the interaction of electron neutrinos with $Z'$ is negligibly small. Then, 
the most stringent constraints come from the Borexino experiment \cite{Bellini:2011rx} that has 
measured the solar neutrinos.
The $^7$Be neutrinos, which is $\nu_e$, oscillate to $\nu_\mu$ and $\nu_\tau$ on the way to the 
Earth and therefore are scattered by electron via $Z'$ exchange. In \cite{Harnik:2012ni}, the constraint from 
Borexino in a $U(1)_{B-L}$ model was studied. We translate the constraint 
given in \cite{Harnik:2012ni}\footnote{The constraint including the interference effect was studied in \cite{Bilmis:2015lja}, 
which showed the constraint is improved by about $30$\%. This effect is important and will be included in our next work.}  
using 
\begin{align}
  g_{B-L} &> \left[(\epsilon e \cos\theta_W)^2 \sum_{j=1}^3f_i|g_{ij}|^2\right]^{1/4}, \\
 |g_{ij}| & \equiv |g' (V^\dagger Q V)_{ij}| = g'
  \begin{pmatrix}
    0.051 & 0.158 & 0.556\\
    0.158 & 0.082 & 0.808\\
    0.556 & 0.808 & 0.133
  \end{pmatrix}, 
\end{align}
where $V$ is the lepton mixing matrix \cite{Maki:1962mu, Pontecorvo:1967fh} 
and $Q = \mathrm{diag}(0,1,-1)$ is the $L_\mu - L_\tau$ charge matrix, $f_i$ stands 
for the fraction of $i$-th mass eigenstate of $^7$Be 
neutrinos at the Earth \cite{Nunokawa:2006ms}. Here, we assumed the normal hierarchy of neutrino 
mass spectrum \cite{Olive:2016xmw}\footnote{This constraint is almost the same for the inverted hierarchy case.}.

\subsection{Beam dump experiment}
Dark photon searches at electron beam dump experiments, such as E141 \cite{Riordan:1987aw} and U70 \cite{Blumlein:2011mv} 
also restrict the model parameters. The coupling constant and kinetic mixing parameter are allowed when the $Z'$ 
boson decays in a beam dump before it reaches to a detector or it is long-lived so that it penetrates a 
detector. The latter case corresponds to too small coupling constant and kinetic mixing which can not explain 
$(g-2)_\mu$. Therefore we consider the former case. The constraint can be translated from the study in the dark photon 
scenario \cite{Essig:2013lka} by
\begin{align}
 \frac{|\epsilon \cos\theta_W|}
       {\sqrt{{\rm Br}(Z'\rightarrow e^+ e^-)}} \gsim \epsilon_{\mathrm{BD}}, \label{eq:beam-dump}
\end{align}
where $\epsilon_{\mathrm{BD}}$ is the kinetic mixing parameter for the dark photon given in \cite{Essig:2013lka}.

\subsection{Meson decay experiment}
Another dark photon searches were performed at the NA48/2 \cite{Batley:2015lha} and E787, E949 \cite{Adler:2004hp, Artamonov:2008qb} 
experiments in which the signals of the dark photon production were searched from the decays of pion and kaon, respectively. 
There analyzed the dark photon decay into an electron and a positron in NA48/2 while the decay into invisible particles in E787/E949. 
The constraints from these experiments give similar bounds and hence we employ the NA48/2 result\footnote{The NA64 collaboration 
recently reported the result of the dark photon search via invisible decays \cite{Banerjee:2016tad}. This result is similar to 
that from the BaBar experiment, and hence we do not consider in this paper. }. 
Then, the constraint can be translated using 
\begin{align}
 |\epsilon \cos\theta_W| \sqrt{{\mathrm{Br}}(Z'\rightarrow e^+ e^-)}
  \lsim
 \epsilon_{\mathrm{MD}},
\end{align}
where $\epsilon_{\mathrm{MD}}$ is the kinetic parameters given in \cite{Batley:2015lha}.

\subsection{Electron-Positron Collider Experiment}
The $Z'$ boson can be directly produced in $e^+$- $e^-$ collision via the kinetic mixing. 
The searches for a light gauge boson such as the dark photon have been performed in $e^+$- $e^-$ 
collider \cite{Babusci:2012cr, Babusci:2014sta}, and 
the most stringent bound is set by the BaBar experiment \cite{Lees:2014xha}.
The $Z'$ boson can decay into charged leptons and be detected as $e^+ + e^- \rightarrow \gamma + l^+ + l^-(l=e,\mu)$,
The constraint can be translated using 
\begin{align}
  |\epsilon \cos\theta_W| \sqrt{{\rm Br}(Z'\rightarrow l^+l^-)}
  \lsim
  \epsilon_{\rm BaBar},
\end{align}
where $\epsilon_{\rm BaBar}$ is the kinetic mixing parameters in the dark photon given in \cite{Lees:2014xha}. 
Furthermore, the constraint for $m_{Z'} > 2 m_\mu$ in the $L_\mu - L_\tau$ model without the kinetic mixing 
was reported in \cite{TheBABAR:2016rlg} by searching the decay of $Z'$ into muons.

\subsection{Electron anomalous magnetic moment}
The $Z'$ boson also contributes to the magnetic moment 
of the electron at the one-loop level similar to the muon. The contribution can be obtained by simply setting  
$g'  = 0$ and replacing $m_\mu$ with the electron mass in Eq.~\eqref{eq:delta-mu-g-2}. 
We require that the $Z'$ contribution to the electron 
magnetic moment $(g-2)_e$ should be within $3\sigma$ \cite{Giudice:2012ms, Aoyama:2014sxa},
\begin{align}
\Delta a_e \lsim 13.8 \times 10^{-13}.
\end{align}

\section{Allowed parameter region} \label{sec:allowed-region}
In this section, we show the allowed region of the parameter space in the $g'$-$\epsilon$ plane taking into account 
the constraints and requirements explained in Sec.~\ref{sec:constraints}. Since the constraints and requirements 
depend on $m_{Z'}$, we choose $m_{Z'} = 10,~50,~100$ and $300$ MeV as illustrating examples.

\begin{figure}[t]
\begin{center}
\begin{tabular}{cc}
\includegraphics[width=8.5cm]{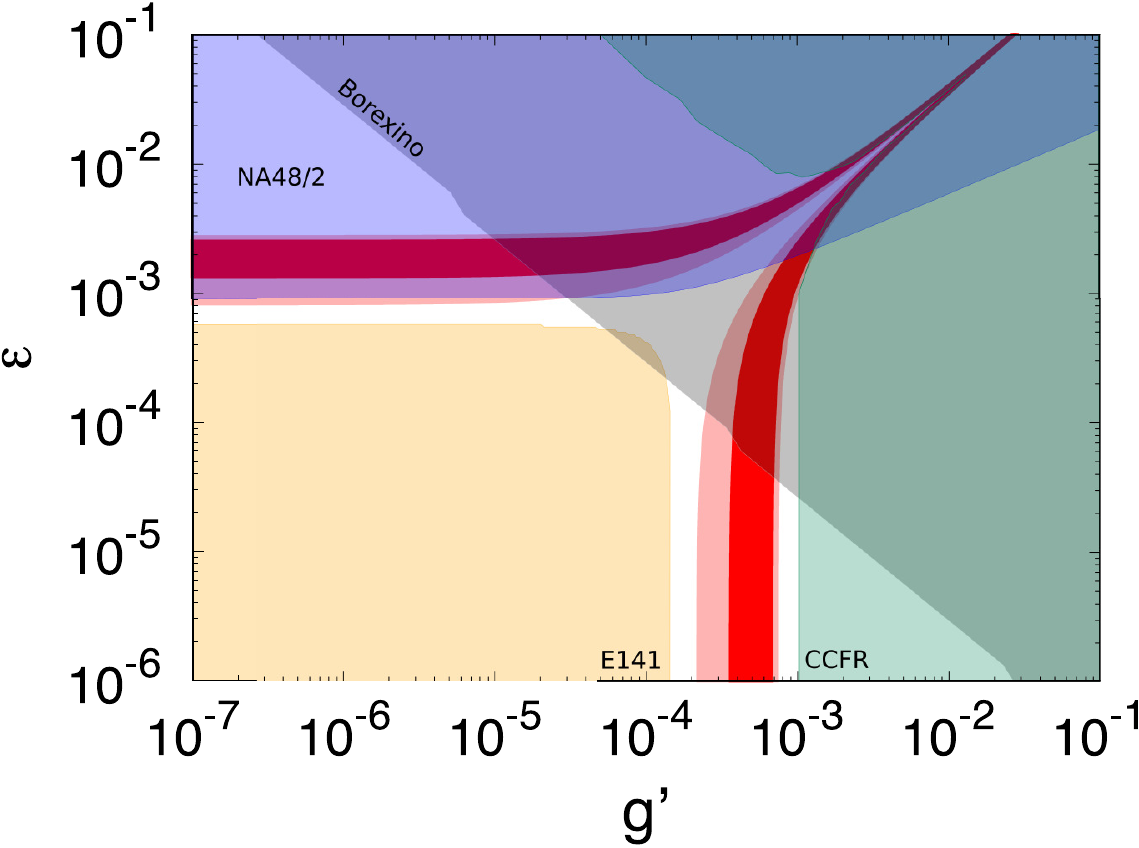}
&
\includegraphics[width=8.5cm]{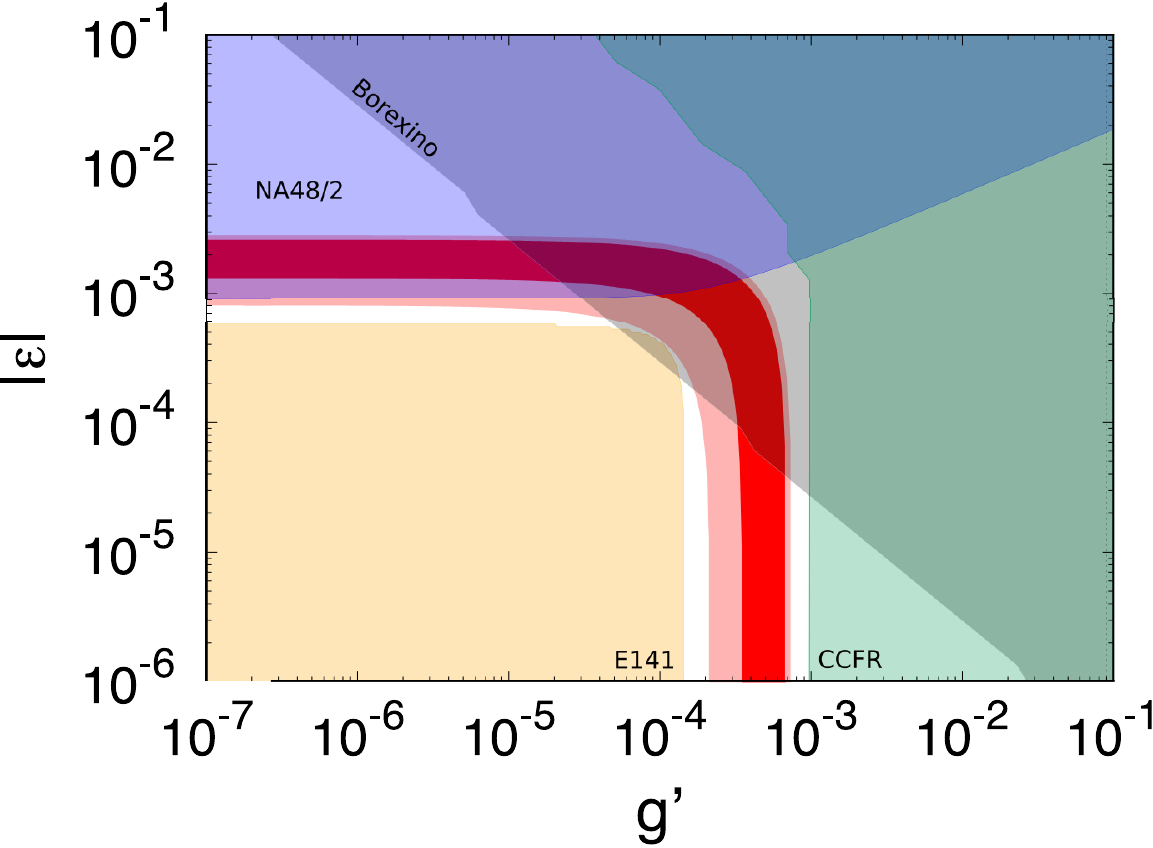} \\
\includegraphics[width=8.5cm]{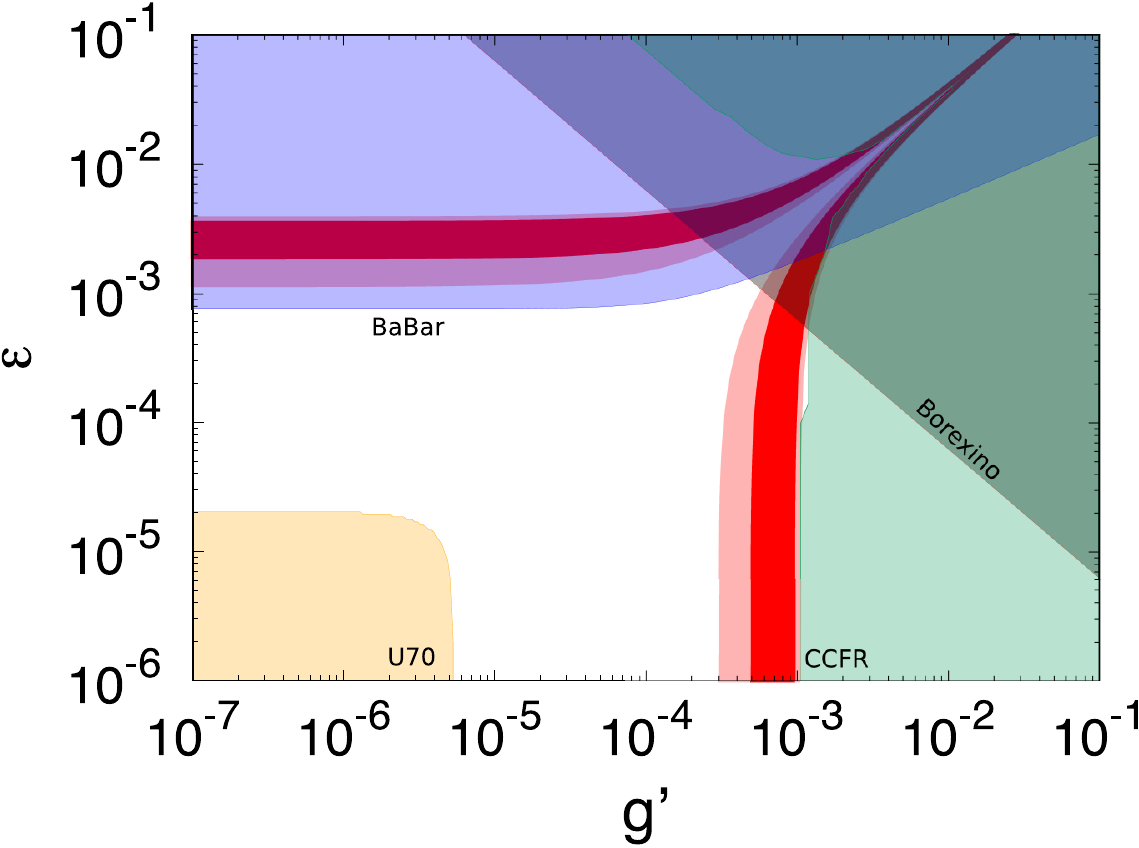}
&
\includegraphics[width=8.5cm]{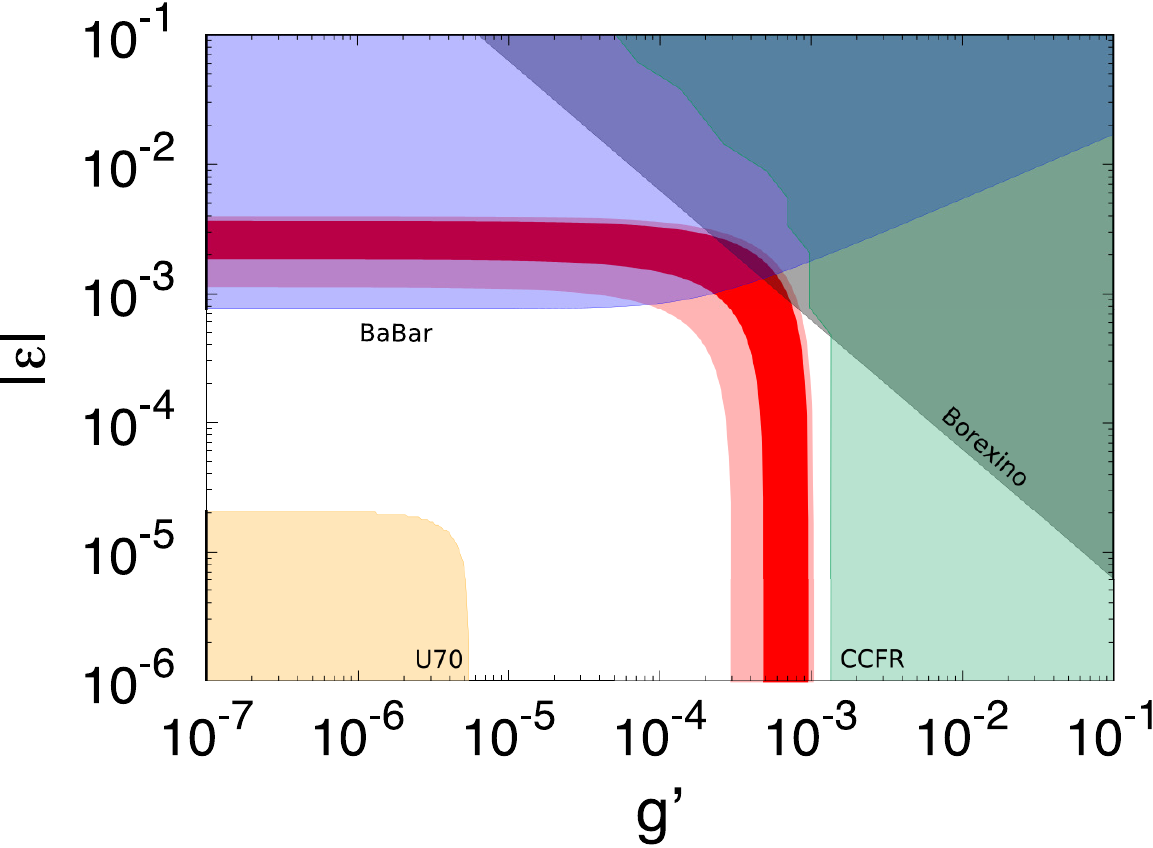} 
\end{tabular}
\end{center}
\caption{ 
The allowed region in the $g'$-$\epsilon$ plane. In the top and bottom panels, $m_{Z'}$ is taken as $10$ and $50$ MeV, 
respectively, and in the left and right panels, $\epsilon$ is positive and negative. The colored regions are excluded 
by the E141/U70 (yellow), the Borexino (grey), the CCFR (green) and $(g-2)_e$ and/or the BaBar (blue) experiments. 
The red and pink bands correspond to $2\sigma$ and $3\sigma$ favored regions of $(g-2)_\mu$.
}
\label{fig:allowed-region1}
\end{figure}

\begin{figure}[t]
\begin{center}
\begin{tabular}{cc}
\includegraphics[width=8.5cm]{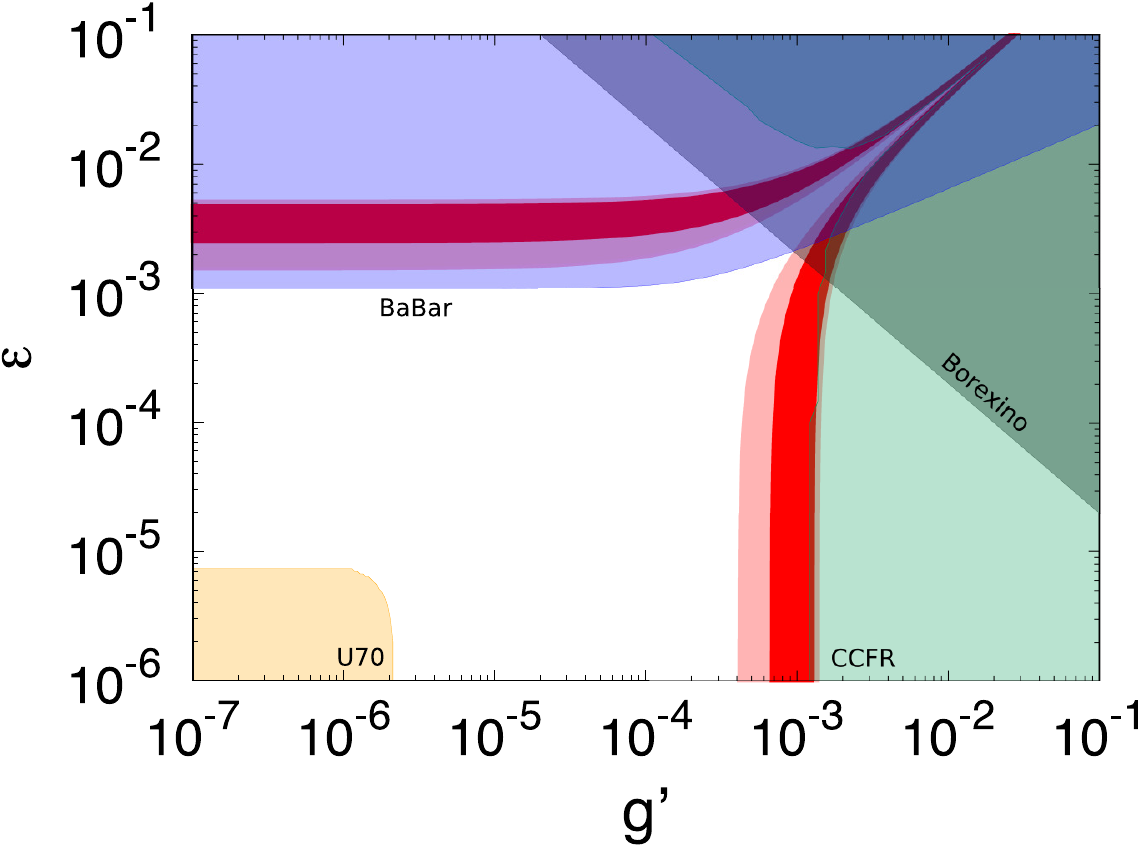}
&
\includegraphics[width=8.5cm]{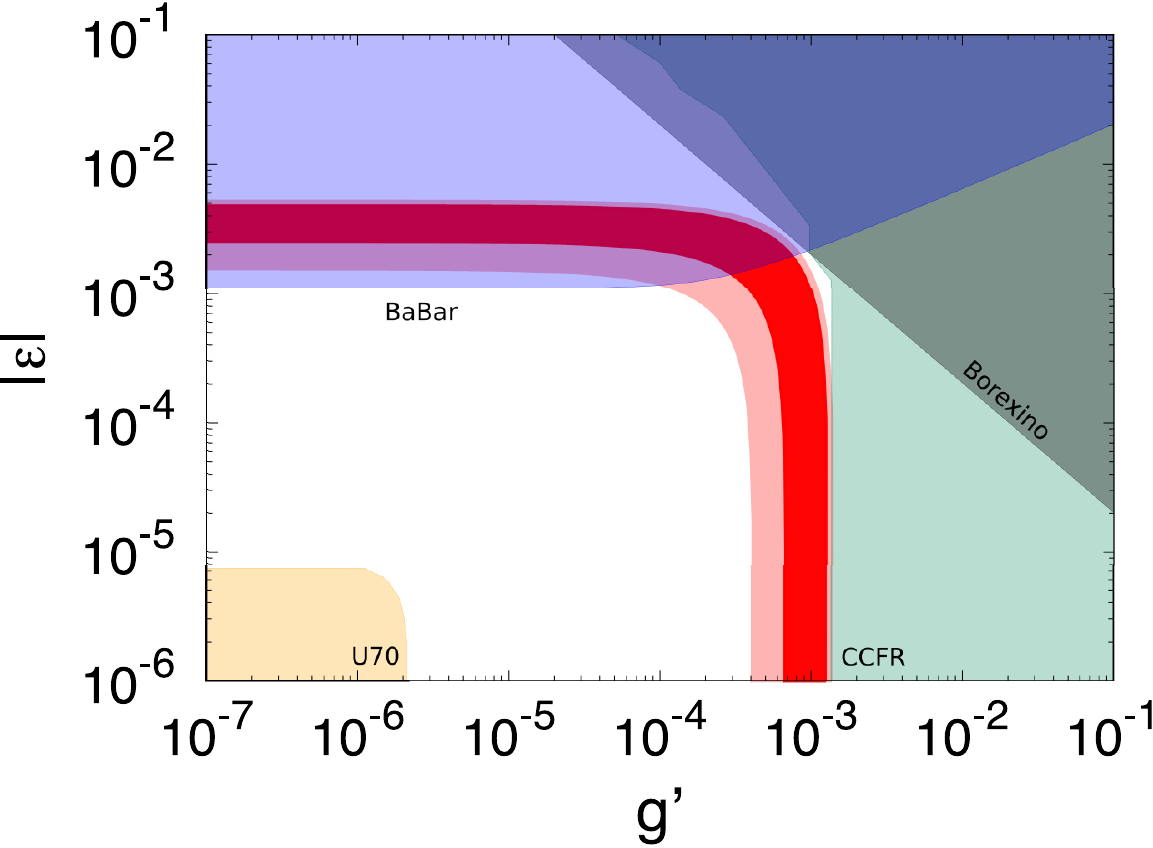} \\
\includegraphics[width=8.5cm]{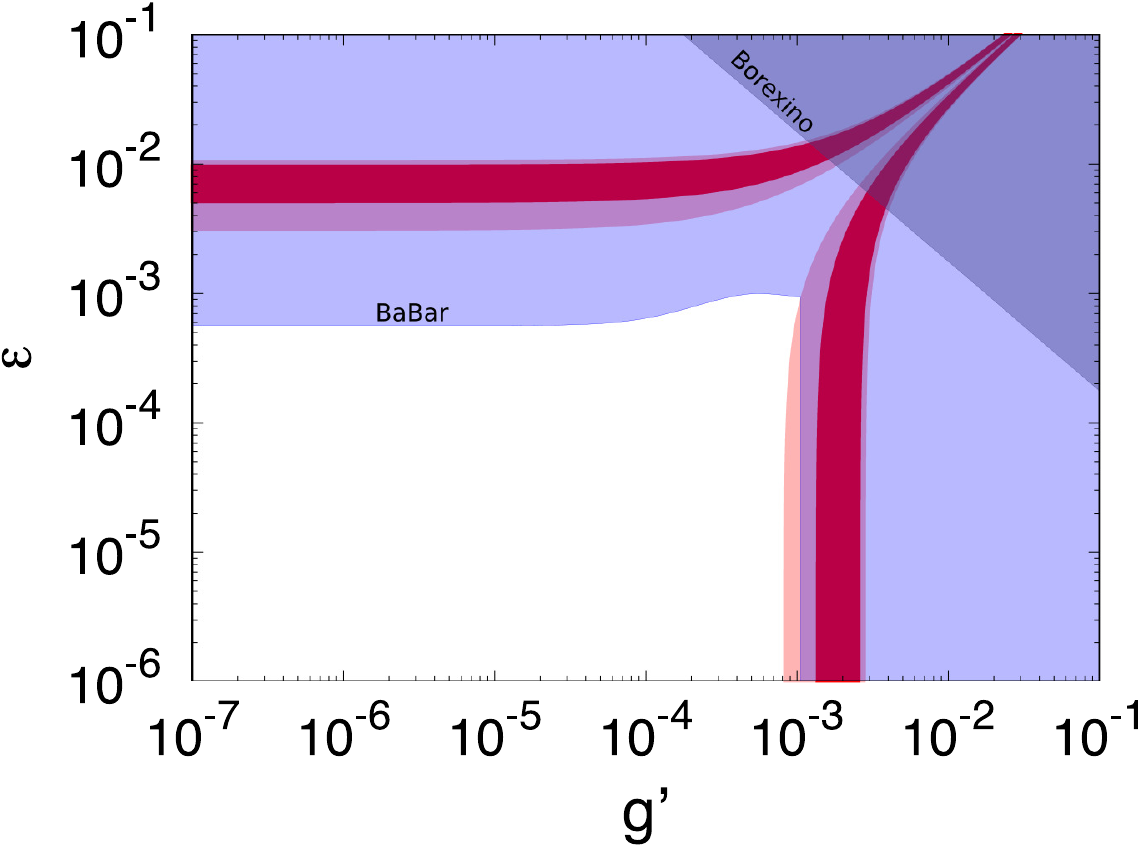}
&
\includegraphics[width=8.5cm]{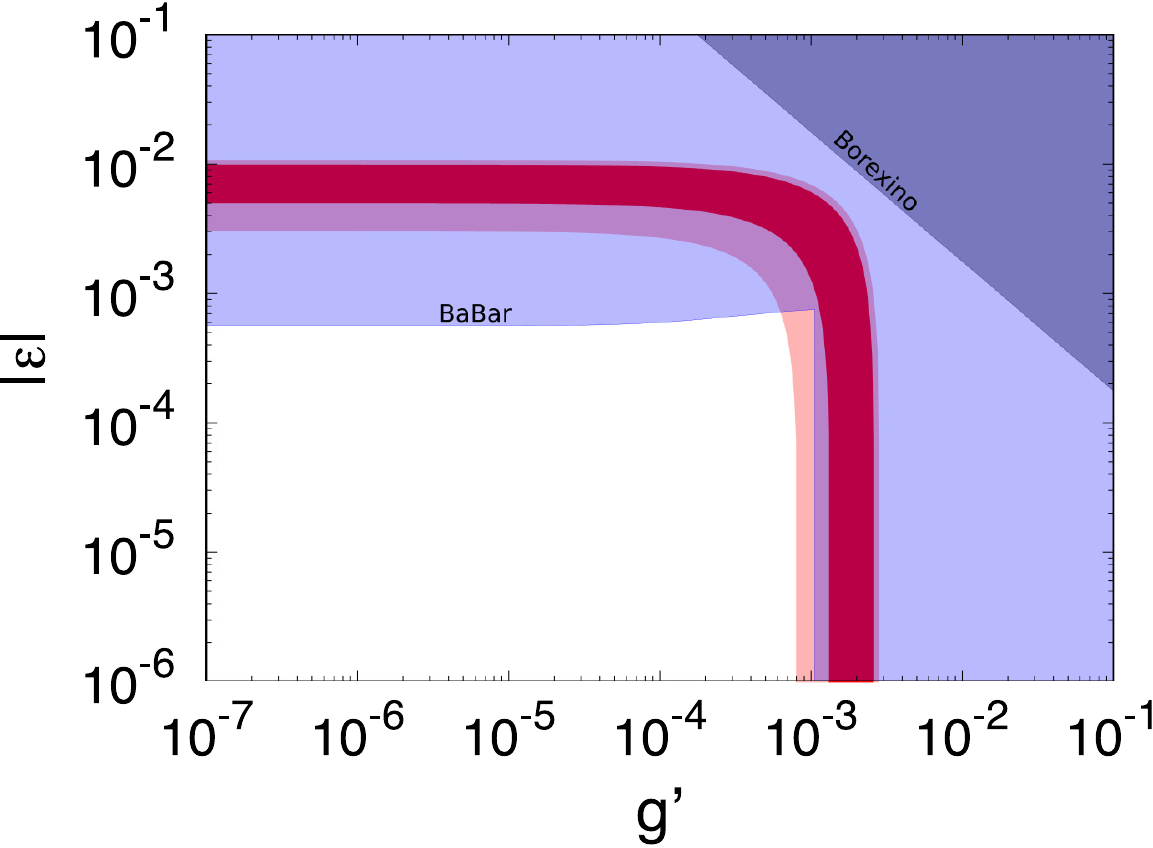} 
\end{tabular}
\end{center}
\caption{ 
The same plots as Figure~\ref{fig:allowed-region1} for $m_{Z'}=100$ (top) and $300$ (bottom) MeV.
}
\label{fig:allowed-region2}
\end{figure}

Figure~\ref{fig:allowed-region1} shows the allowed region in the $g'$-$\epsilon$ plane. The mass of $Z'$ is taken 
as $10$ MeV and $50$ MeV for the top and bottom panels, and the kinetic mixing parameter is taken to be positive and negative 
for the left and right panels, respectively. 
In the figure, the yellow, grey and green regions are excluded by the E141/U70 (beam dump), 
the Borexino ($\nu$-$e$ scattering) and the CCFR (neutrino trident production) experiments. The blue region is 
also excluded by $(g-2)_e$ and/or the BaBar ($e^+$- $e^-$ collider) and/or the NA48/2 (meson decay) experiment.  
The red and pink bands represent the favored regions 
of $(g-2)_\mu$ within $2\sigma$ and $3\sigma$, respectively. Figure~\ref{fig:allowed-region2} is the same plots 
for $m_{Z'}=100$ and $300$ MeV. 

From these figures, one can see that the $(g-2)_\mu$ favored regions are different with the sign of $\epsilon$. 
In the case of positive $\epsilon$ (left panels), the favored region of $(g-2)_\mu$ is extended to the right-upper corner. 
This is because the coupling of the muon becomes smaller due to the 
cancellation between $g'$ and $\epsilon$. In this region, the constraint from the CCFR experiment can be evaded.  
Then, slightly larger values of $g'$ is allowed for $m_{Z'} = 100$ MeV. 
In the case of negative $\epsilon$ (right panels), on the other hand, the coupling becomes larger due to the addition of 
$g'$ and $\epsilon$. 
One can also see that the constraint from CCFR is more stringent in negative $\epsilon$ than in positive $\epsilon$ 
for $|\epsilon| \gg g'$.  In this parameter region, the coupling of the muon is given by $- \epsilon e \cos\theta_W$, and 
therefore the relative phase of the amplitudes for the neutrino trident process is determined 
by the sign of $\epsilon$. Then, the amplitudes are added destructively for positive $\epsilon$ while 
constructively for negative $\epsilon$. The difference of  the excluded region by CCFR comes from this fact.

It is seen that, for $g' \lsim 10^{-4}$,  the BaBar or the NA48/2 experiments excludes the 
regions with roughly $|\epsilon| \gsim 10^{-3}$ for $m_{Z'} \lsim 100$ MeV and $|\epsilon| \gsim 5 \times 10^{-4}$ 
for $m_{Z'} = 300$ MeV, respectively.
Therefore, $(g-2)_\mu$ can not be explained within $3\sigma$ with $g' \lsim 10^{-4}$ for $m_{Z'} \gsim 50$ MeV 
in our example parameters.  
This result generally holds for different values of $m_{Z'}$ because such the small $g'$ does not change 
the constraint given in \cite{Lees:2014xha, Batley:2015lha}.
On the other hand, for $m_{Z'} = 10$ MeV, the allowed region including $(g-2)_\mu$ within $3\sigma$ is found. 
One will find similar allowed regions for some values of $m_{Z'} \lsim 20$ MeV because the constraint from \cite{Batley:2015lha} 
becomes less stringent due to statistical fluctuations. 

For $|\epsilon| \lsim 10^{-3}$, it is seen that the regions with roughly $g' \gsim 10^{-3}$ are excluded by 
the CCFR experiment for $m_{Z'} \lsim 100$ MeV and the BaBar experiment for $m_{Z'} = 300$ MeV. 
For $m_{Z'} = 10$ MeV, the E141 experiment also has excluded for 
$g' \lsim 1.3 \times 10^{-4}$, and the Borexino has set the upper limit on $\epsilon \lsim 2 \times 10^{-4}$. 
Then, the parameter space is much constrained, however $(g-2)_\mu$ within $3\sigma$ is still allowed.
For $m_{Z'} \gsim 50$ MeV, the constraints from the beam dump experiments become weaker.
These constraints come from that $Z'$ is short lived so that it decays before reaching to a detector, 
as we mentioned in Sec.~\ref{sec:constraints}. 
Since the lifetime is inversely proportional to the coupling constant squared times $m_{Z'}$, the coupling constant 
can be smaller as $m_{Z'}$ is larger. This $m_{Z'}$ dependence is incorporated in the 
values of $\epsilon_{\mathrm{BD}}$ given in \cite{Essig:2013lka} . 
The $g'$ and $\epsilon$ dependences of the excluded region can be understood by Eq.~\eqref{eq:beam-dump}.
For $m_{Z'} = 300$ MeV, we superposed the constraint on $g'$  (the vertical line) read from \cite{TheBABAR:2016rlg}. 
Strictly speaking, the constraint depends on $\epsilon$. However it may not be so different because $g'$ is larger 
than $\epsilon$ in this region.

\section{Light $Z'$ search at Belle-II and neutrino beam experiments} \label{sec:results}

Based on the results shown in Sec.~\ref{sec:allowed-region}, we study the possibilities on the search for the $Z'$ boson at 
the Belle-II and neutrino beam experiments.

\subsection{The Belle-II experiment}

\begin{figure}[t]
\begin{center}
\includegraphics[width=10cm]{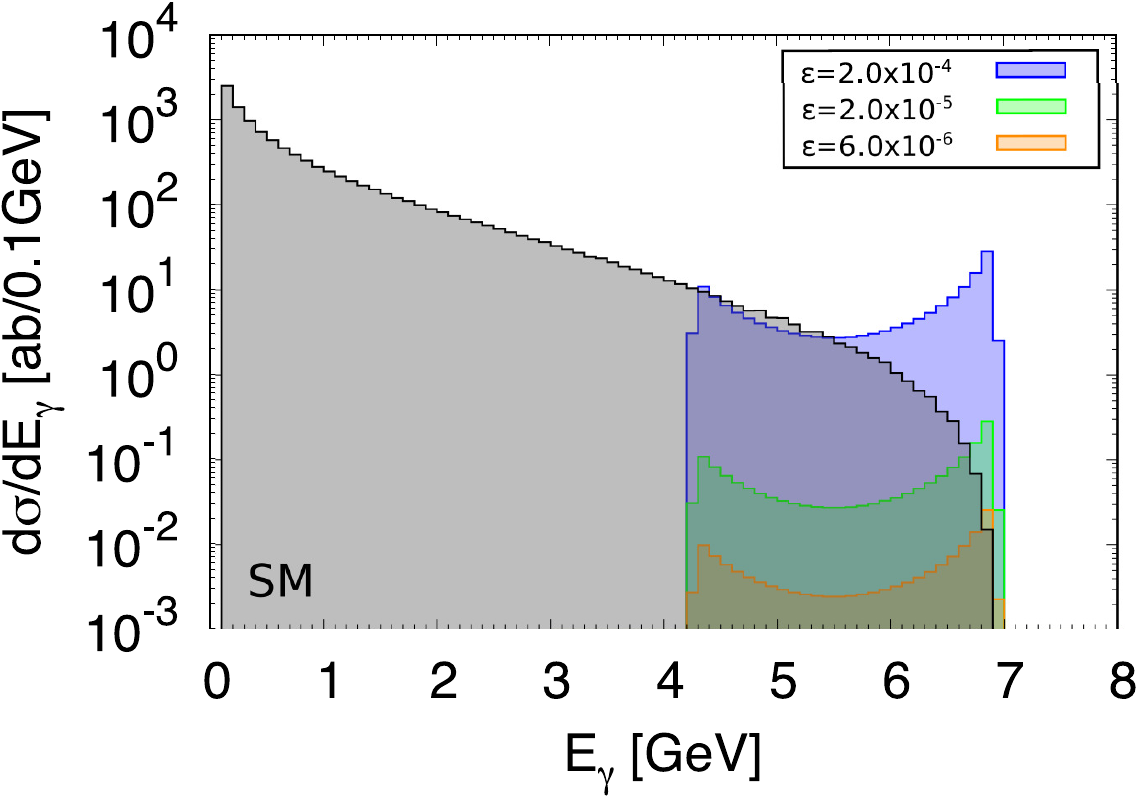}
\end{center}
\caption{ 
The differential cross section of $e^+ + e^- \rightarrow \gamma + Z'$ with respect to the photon energy, 
$E_\gamma$. The blue, green and the orange histograms correspond to $\epsilon = 2 \times 10^{-4},~2 \times 10^{-5}$ 
and $6 \times 10^{-6}$, respectively. The grey histogram represents the SM background.
}
\label{fig:diff-cross-Egamma}
\end{figure}

The Belle-II experiment is an $e^+ e^-$ collider at the center of mass energy $\sqrt{s}=10.58$ GeV \cite{Abe:2010gxa}. 
Its goal is to accumulate the integrated luminosity $50$ ab$^{-1}$ of $e^+ e^-$ collision data during by the middle of 
the next decade.
In $e^+ e^-$ collision, the $Z'$ boson can be produced through the kinetic mixing \cite{Ma:1978zm, Carena:2003aj, Fayet:2007ua}, 
and then decays into neutrinos, charged leptons and pions. 
The processes with charged leptons and pions in the final states will be overwhelmed by the SM backgrounds because those can 
occur by the electromagnetic interaction.  
The process with neutrinos, on the other hand, occurs by the weak interaction in the SM, and it is suppressed by 
the $W$ and $Z$ boson mass. 
Therefore the signal can be comparable to or larger than the background. 
Furthermore, the signal of the $Z'$ production can be characterized by the energy  of an associated photon.\footnote{A similar 
searched was done at BaBar for a pseudo scalar \cite{Aubert:2008as}.} The energy 
of the photon is given by
\begin{align}
E_\gamma = \frac{4 E_{e^+} E_{e^-} - m_{Z'}^2}{2( E_{e^+} + E_{e^-} + (E_{e^+} - E_{e^-} ) \cos\theta_\gamma)},
\end{align}
where $E_{e^\pm}$ is the energy of positron and electron, and $\theta_\gamma$ is the angle between the photon momentum and 
the electron momentum. 
Here we ignored the angle between the positron and electron momenta for simplicity.
The Belle-II detector can identify the photon for $E_\gamma \geq 0.1$ GeV with the resolution $0.1$ GeV and 
the angle $15^\circ \leq \theta_\gamma \leq 135^\circ$ \cite{Abe:2010gxa}. 
With these cuts, the photon energy ranges $4.3 \leq E_\gamma \leq 6.9$ GeV. 

Figure \ref{fig:diff-cross-Egamma} shows the differential cross section of $e^+ + e^- \rightarrow \gamma + Z'$ 
with respect to the photon energy, $E_\gamma$.  
The blue, green and the orange histograms correspond to $\epsilon = 2 \times 10^{-4},~2 \times 10^{-5}$ 
and $6 \times 10^{-6}$, respectively. The grey histogram represents the SM background of $\gamma+$ missing events, 
which comes from $e^+ + e^- \rightarrow \gamma + Z^\ast \rightarrow \gamma + \nu + \bar{\nu}$ and also t-channel $W$ 
exchange one. 
The mass of $Z'$ is fixed to $100$ MeV, however the differential cross section is almost independent of the mass 
for $m_{Z'} \lsim 300$ MeV.
It can be seen from the figure that the differential cross section of the $Z'$ production is different from the SM background. 
The deviations from the background become significant as $\epsilon$ becomes larger.  
The expected numbers of events in the last 
two bins are $1500$, $15$ and $1.4$ for each $\epsilon$, respectively while that of the SM background is less than $1$. 
Therefore the search for $Z'$ will be possible even for $\epsilon = 6 \times 10^{-6}$ by measuring 
the mono photon events with the $E_\gamma \gsim 6.8$ GeV. 
%

\begin{figure}[t]
\begin{center}
\begin{tabular}{cc}
\includegraphics[width=8.5cm]{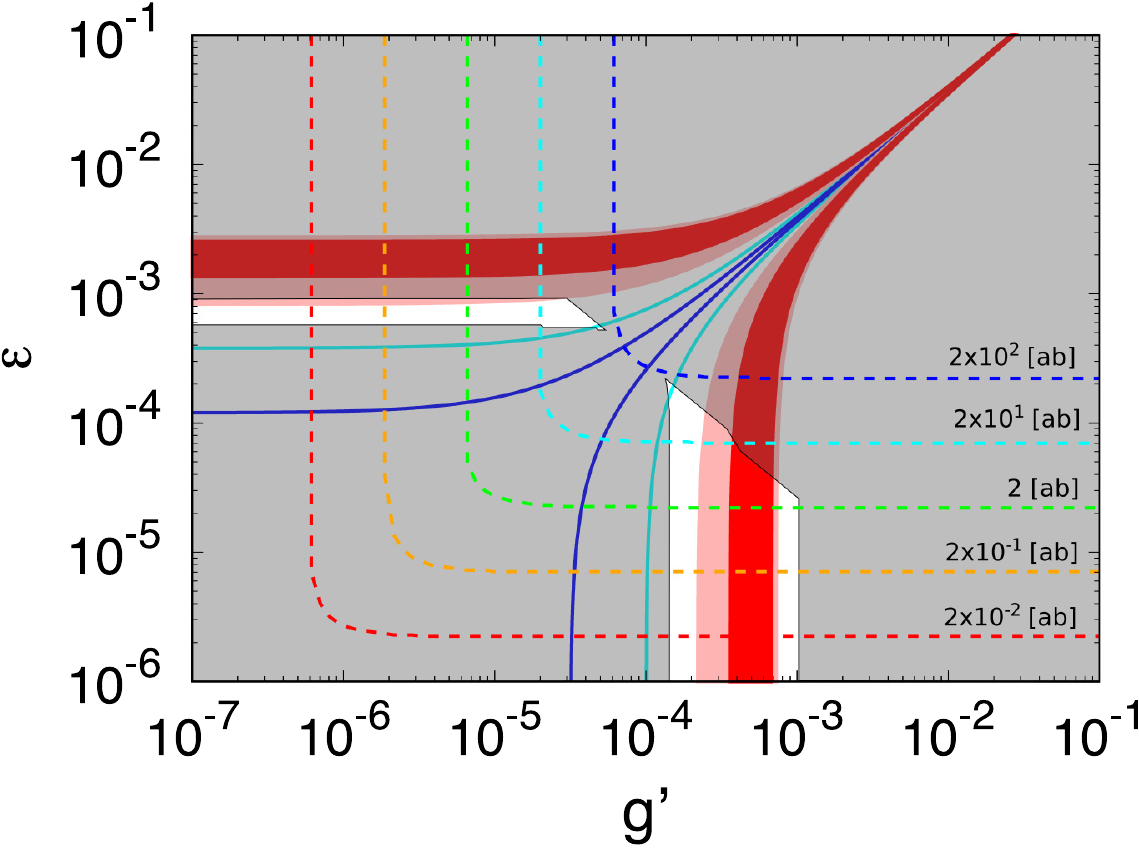}
&
\includegraphics[width=8.5cm]{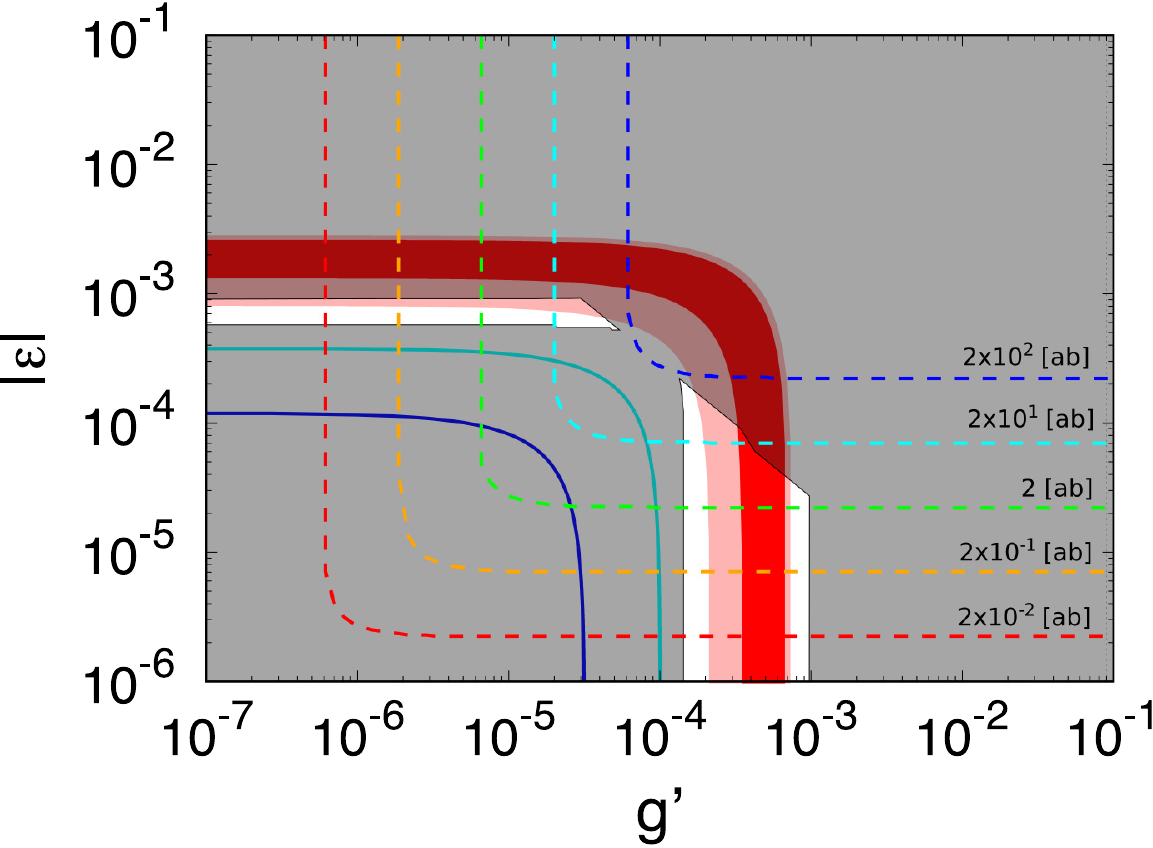} \\
\includegraphics[width=8.5cm]{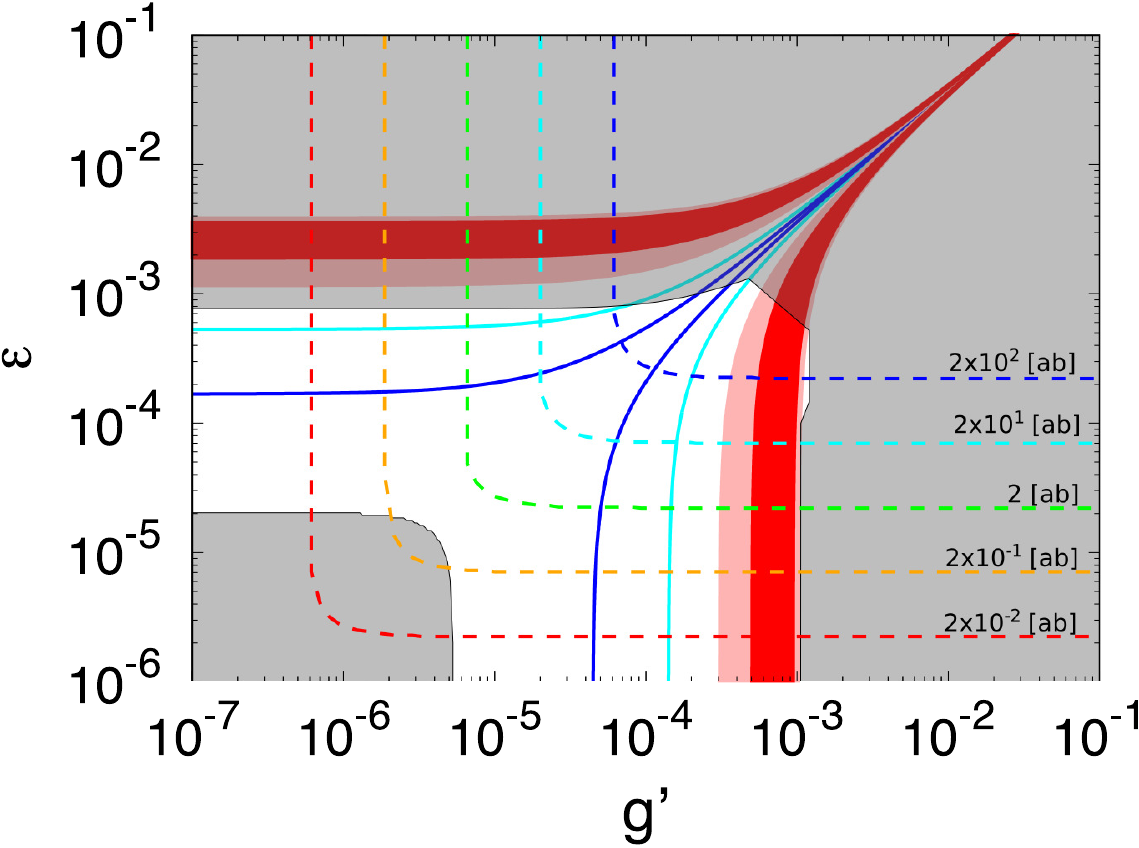}
&
\includegraphics[width=8.5cm]{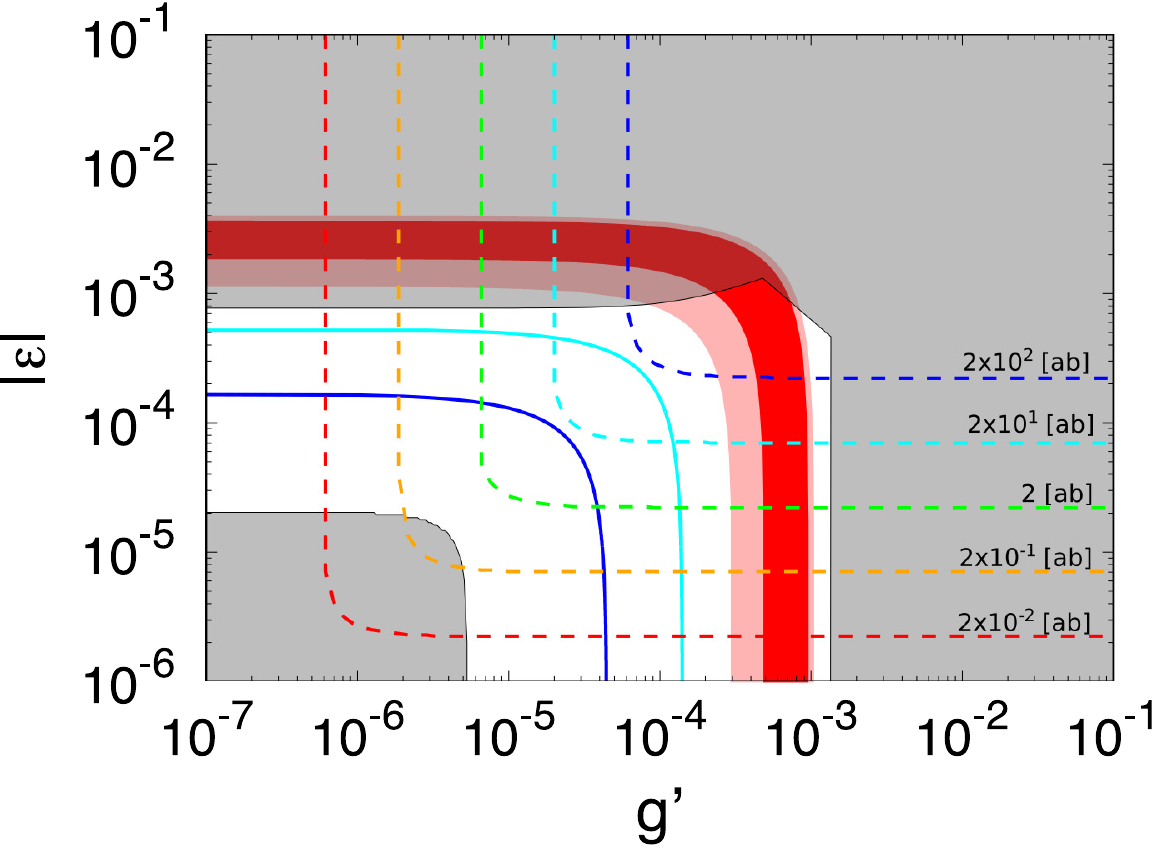} 
\end{tabular}
\end{center}
\caption{The contour plots of the total cross section of $e^+ + e^- \rightarrow \gamma + \nu + \bar{\nu}$ for 
$10$ (top) and $50$ (bottom) MeV. The left and right panels correspond to $\epsilon > 0$ and $\epsilon < 0$ 
cases, respectively. The numbers near each dashed curves are the cross sections in ab. 
The red and pink bands represent $(g-2)_\mu$ within $2\sigma$ and $3\sigma$, and the solid cyan and blue 
curves represent $\Delta a_\mu = 10^{-10}$ and $10^{-11}$, respectively.
}
\label{fig:belle-g-2-1}
\end{figure}

\begin{figure}[t]
\begin{center}
\begin{tabular}{cc}
\includegraphics[width=8.5cm]{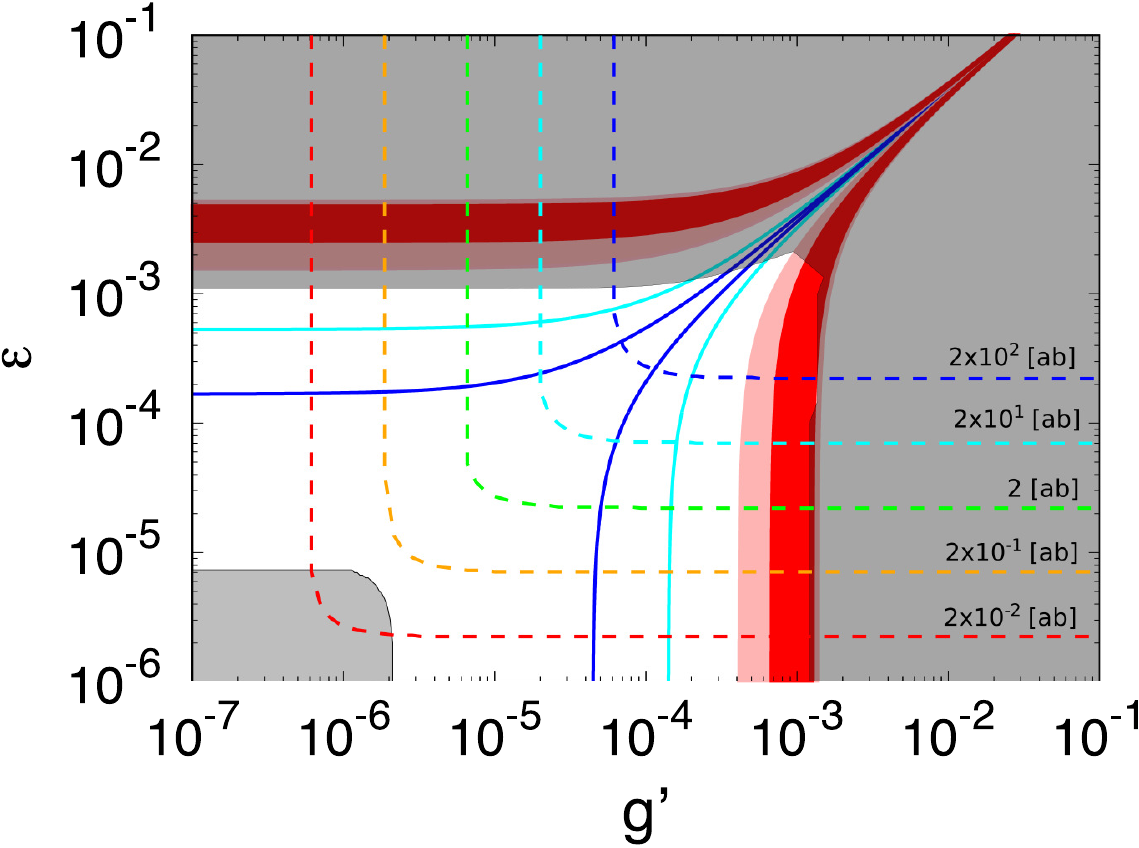}
&
\includegraphics[width=8.5cm]{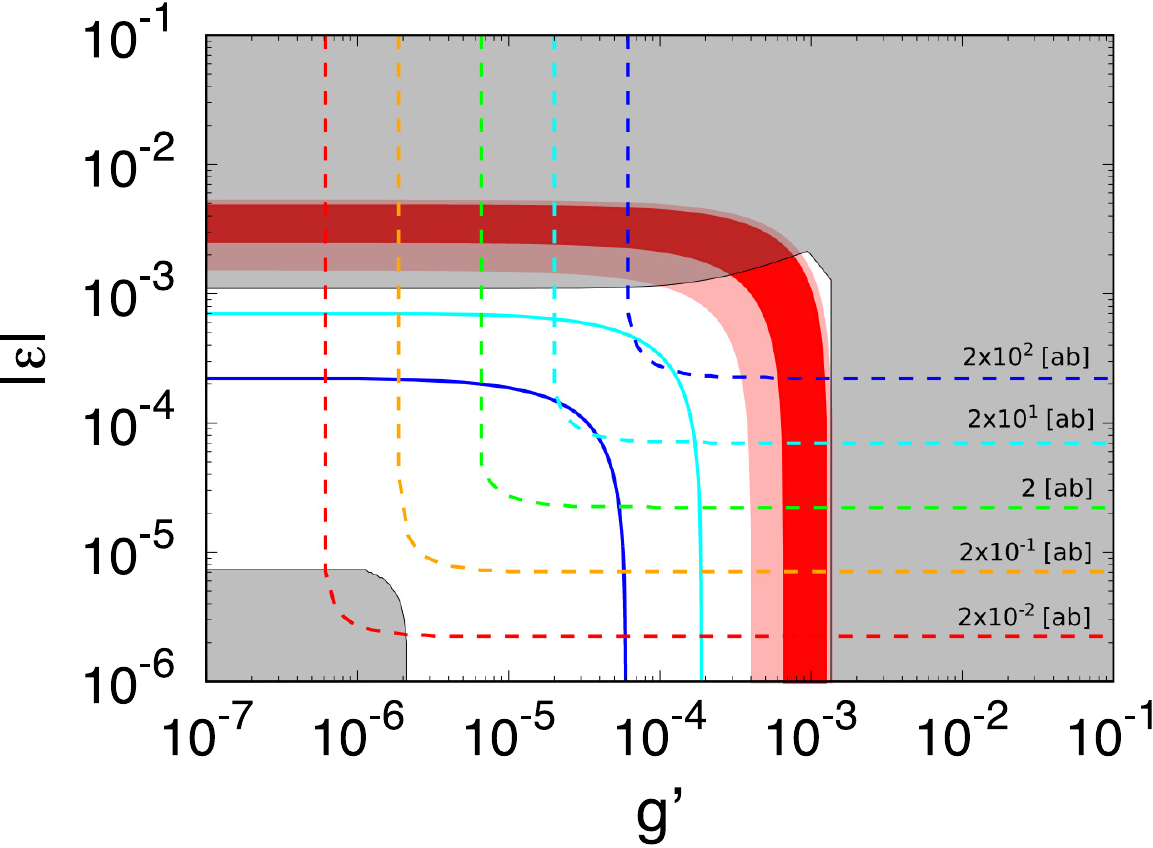} \\
\includegraphics[width=8.5cm]{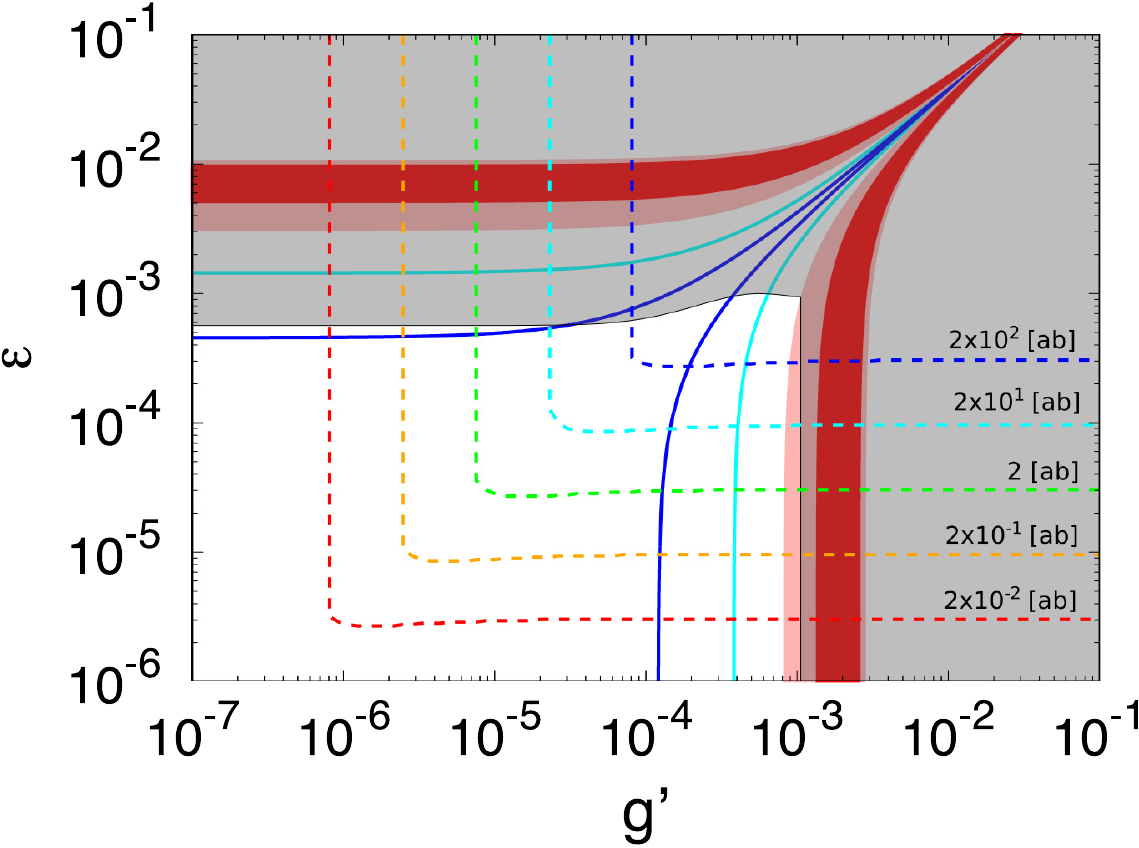}
&
\includegraphics[width=8.5cm]{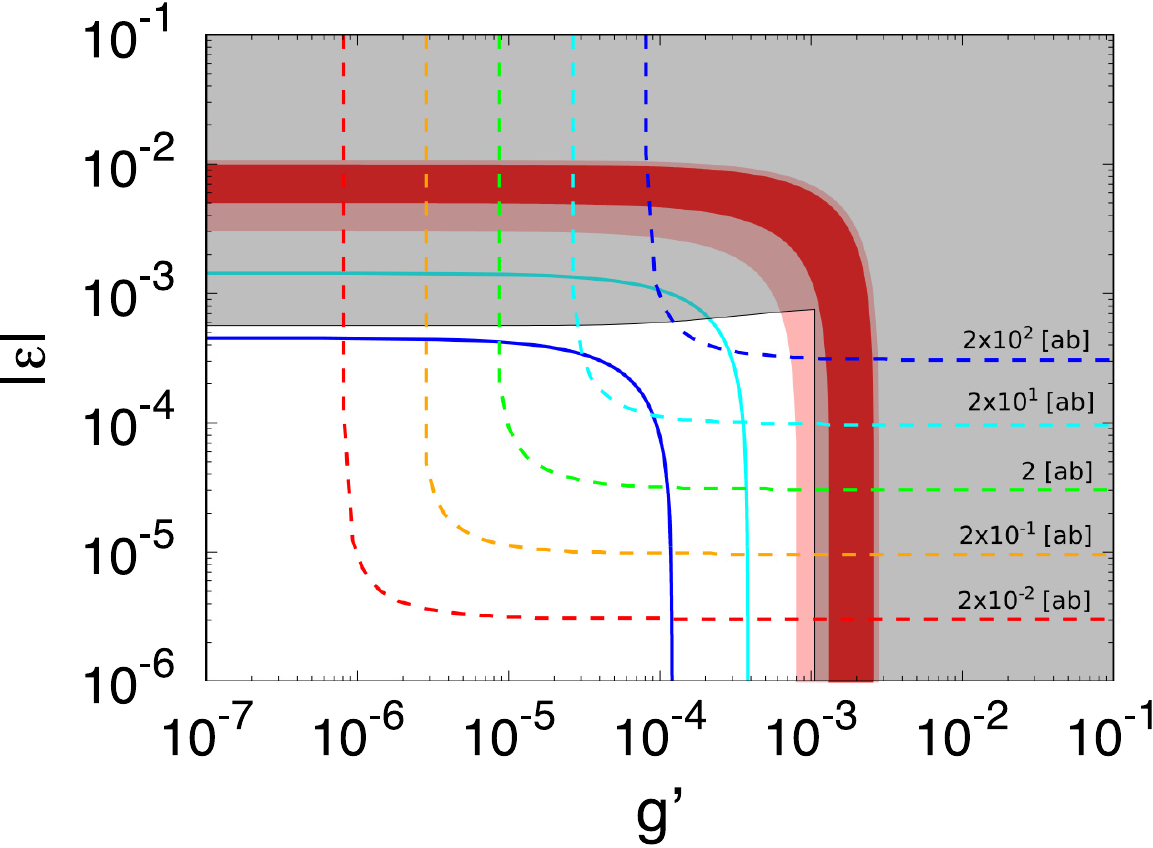} \\
\end{tabular}
\end{center}
\caption{ 
The same plots as Figure \ref{fig:belle-g-2-1} for $m_{Z'} = 100$ (top) and $300$ (bottom) MeV.}
\label{fig:belle-g-2-2}
\end{figure}

Figures~\ref{fig:belle-g-2-1} and \ref{fig:belle-g-2-2} are the contour plots of the cross section of $e^+ + e^- \rightarrow \gamma + Z'$ 
followed by $Z' \rightarrow \nu + \bar{\nu}$ in the $g'$-$\epsilon$ plane, where the decay branching 
ratio of $Z' \rightarrow \nu + \bar{\nu}$ is obtained from Eqs.~\eqref{eq:decay-width}. 
In each panel, the mass of $Z'$ and the sign of $\epsilon$ are the same 
as Figures \ref{fig:allowed-region1} and \ref{fig:allowed-region2}, respectively. 
The dashed curves represent the contours of the cross section between $200$ to $0.02$ ab from top to bottom. 
Assuming the luminosity $50$ ab$^{-1}$, the expected numbers of events to each cross sections are 
from $10^4$ to $1$.
The grey regions are the excluded region in Figs.~\ref{fig:allowed-region1} and \ref{fig:allowed-region2}, and 
the red and pink bands represent the favored regions of $(g-2)_\mu$ within $2\sigma$ and $3\sigma$. 
The solid cyan and blue curves represent $\Delta a_\mu = 10^{-10}$ and $10^{-11}$ for references. 
When the planned experiments reduce the uncertainties and if the same-level progresses on theoretical side are made, 
such smaller contributions to $(g-2)_\mu$ might be required. 

The shape of the contours can be understood as follows. The production cross section of $Z'$ is proportional to 
$\epsilon^2$ while the decay branching ratio is proportional to $g'^2/(g'^2 + \epsilon^2 + \cdots)$. Thus, the total 
cross section is proportional to $\epsilon^2 g'^2/(g'^2 + \epsilon^2 + \cdots)$. When $\epsilon$ is much smaller 
than $g'$, the total cross section is independent of $g'$.  In the opposite situation, $\epsilon \gg g'$, the cross section 
becomes independent of $\epsilon$. 
It is important to be noted here that the differential cross section with respect to $E_\gamma$ is the same on 
each contour even if the branching ratio is different. This is because the shape of the different cross section is 
determined by the production cross 
section and the magnitude of that is determined by the total cross section.

The contour of $0.2$ ab is close to the case of $\epsilon = 6 \times 10^{-6}$ in figure \ref{fig:diff-cross-Egamma}. 
From the figures \ref{fig:belle-g-2-1} and \ref{fig:belle-g-2-2}, it can be seen that the contour of $0.2$ ab 
covers the region of $g' \gsim 2 \times 10^{-6}$ and $\epsilon \gsim 7 \times 10^{-6}$.  
As discussed in Fig.~\ref{fig:diff-cross-Egamma}, the signal is larger than the SM background and hence 
this region will be explored. 
Furthermore, the curves of $\Delta a_\mu = 10^{-10}$ and $10^{-11}$ are covered in this region. Therefore, not 
only the present $(g-2)_\mu$ favored regions but also smaller ones can be examined by the Belle-II experiment.

\subsection{Neutrino Beam Experiments}
Next we discuss the detection possibilities of the $Z'$ boson at neutrino beam experiments through 
the neutrino trident production process\footnote{Some results in this subsection overlap with \cite{Magill:2016hgc} which appeared on 
arXiv while our manuscript had been prepared. The results were presented at 
"The international workshop on future potential of high intensity accelerators for particle and nuclear physics (HINT2016)", at 
J-PARC, Tokai, Japan and other places.}.

\begin{figure}[t]
\begin{center}
\begin{tabular}{cc}
\includegraphics[width=8.5cm]{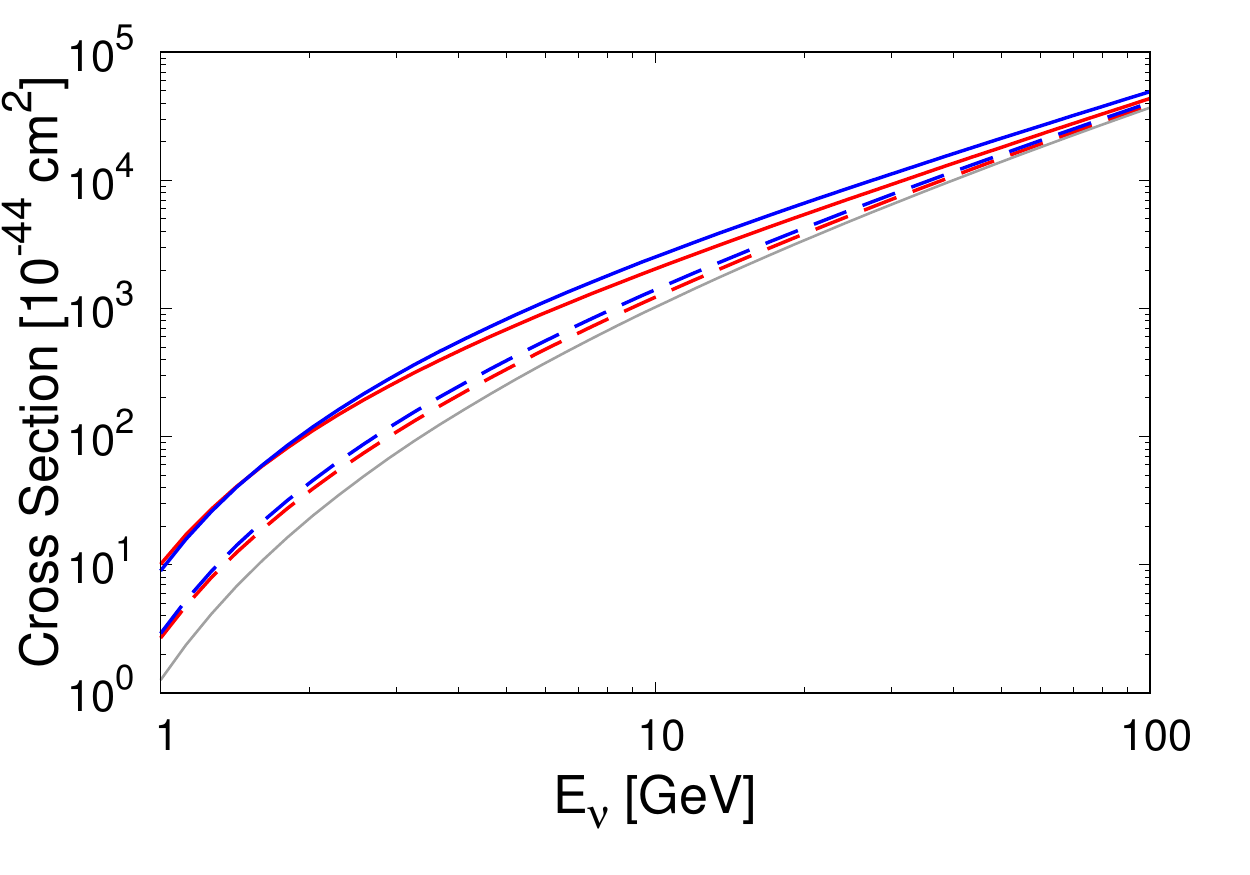}
&
\includegraphics[width=8.5cm]{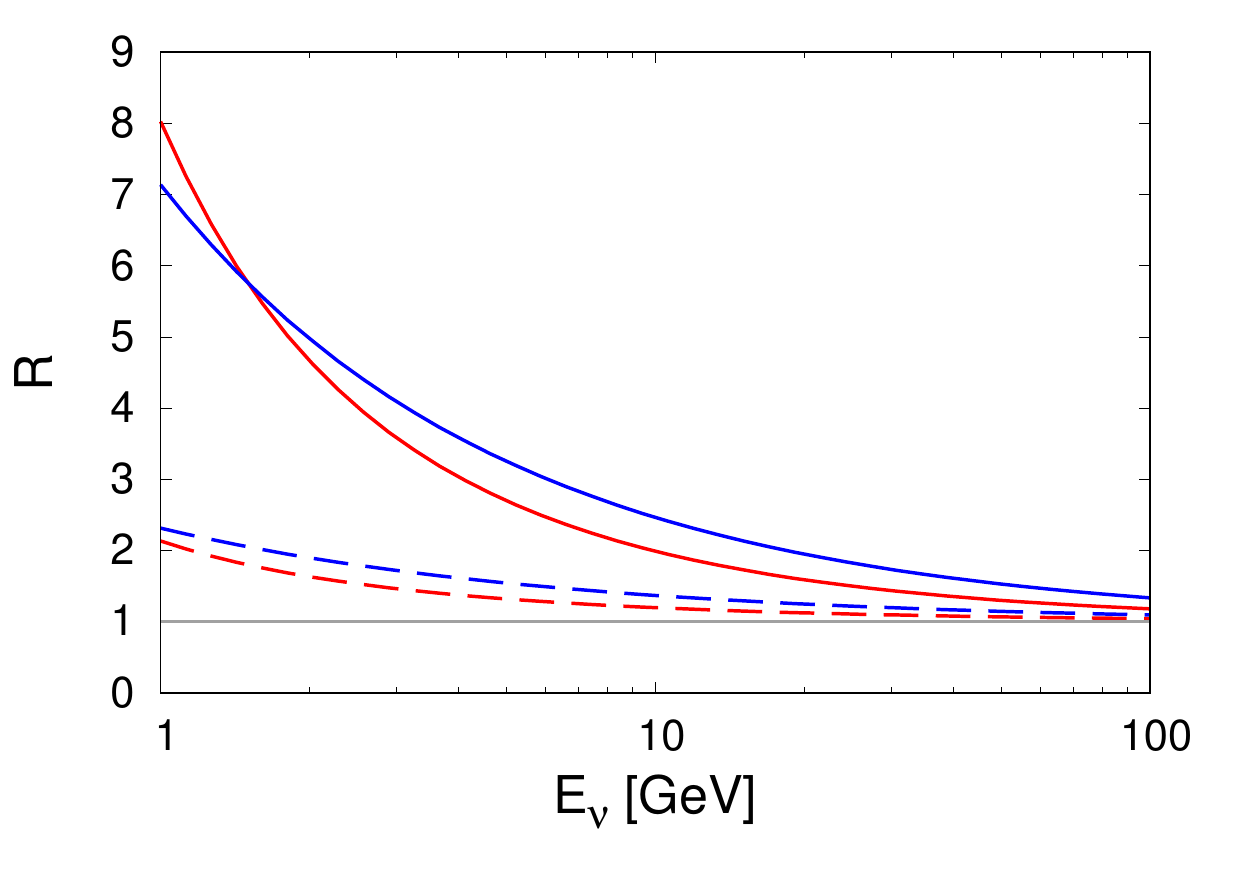} \\
\end{tabular}
\end{center}
\caption{ 
The cross section of the neutrino trident production process (left) and the ratio of the cross section to the 
SM one, $R$,  (right) in terms of the neutrino energy for an iron target.
The kinetic mixing parameter is fixed to $10^{-5}$, and the $Z'$ mass is taken to be $10$ (red) and $100$ (blue) MeV, 
respectively.
The gauge coupling constant is taken as $g'=5.8 \times 10^{-4}$ (red-solid), $3.4 \times 10^{-4}$ (red-dashed), 
and $g'=9.5 \times 10^{-4}$ (blue-solid), $5.8 \times 10^{-4}$ (blue-dashed), respectively. 
The grey curve represents the SM cross section.
}
\label{fig:trident-CS}
\end{figure}

\begin{figure}[t]
\begin{center}
\begin{tabular}{cc}
\includegraphics[width=8.5cm]{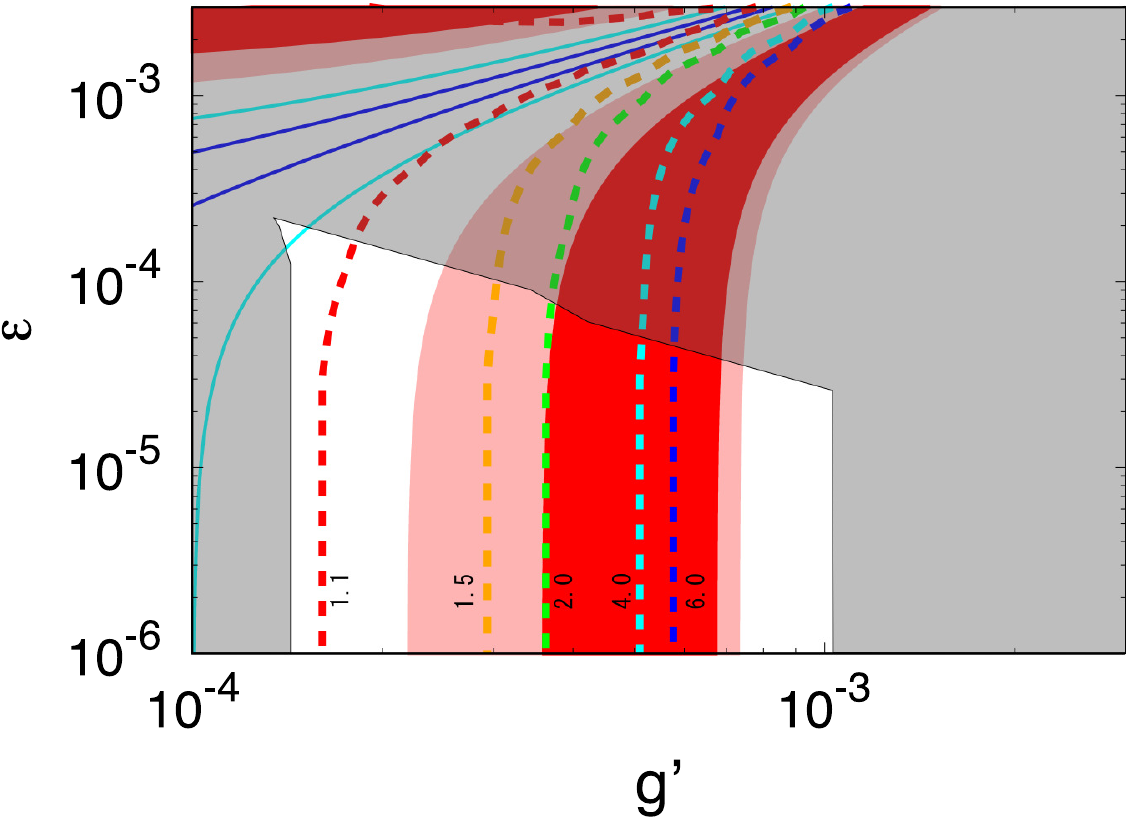}
&
\includegraphics[width=8.5cm]{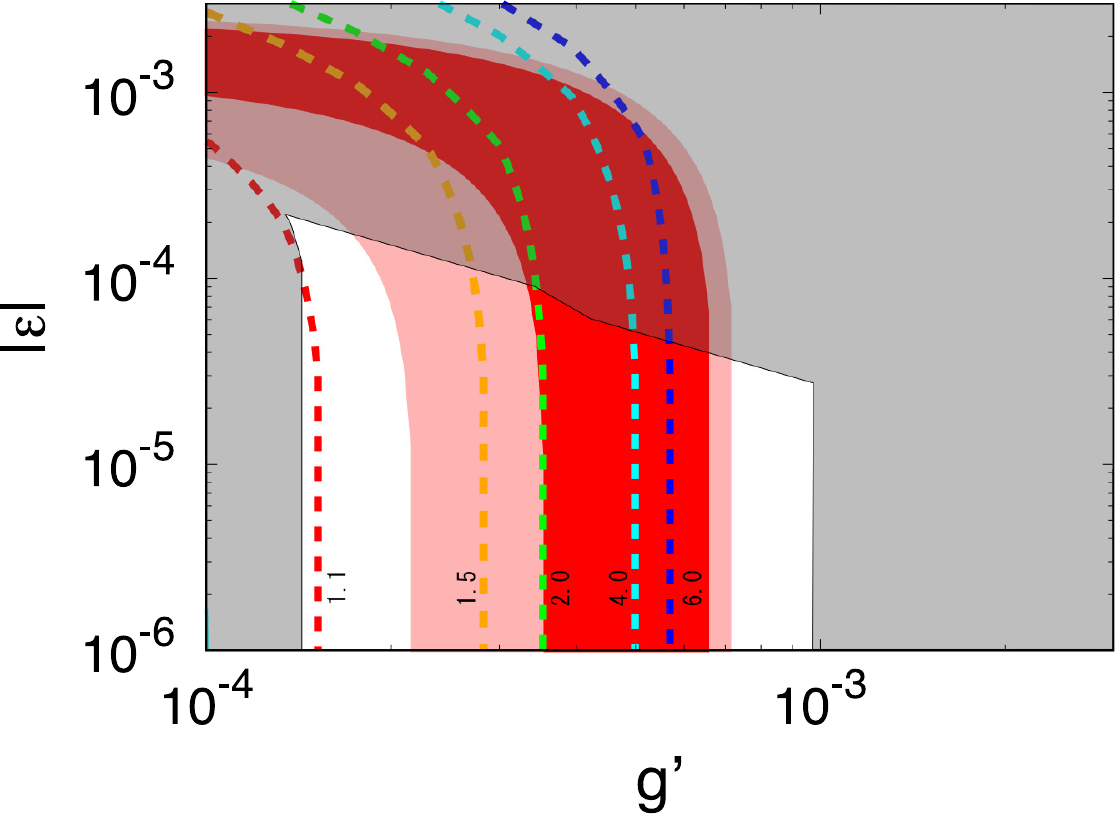} \vspace{2mm} \\
\includegraphics[width=8.5cm]{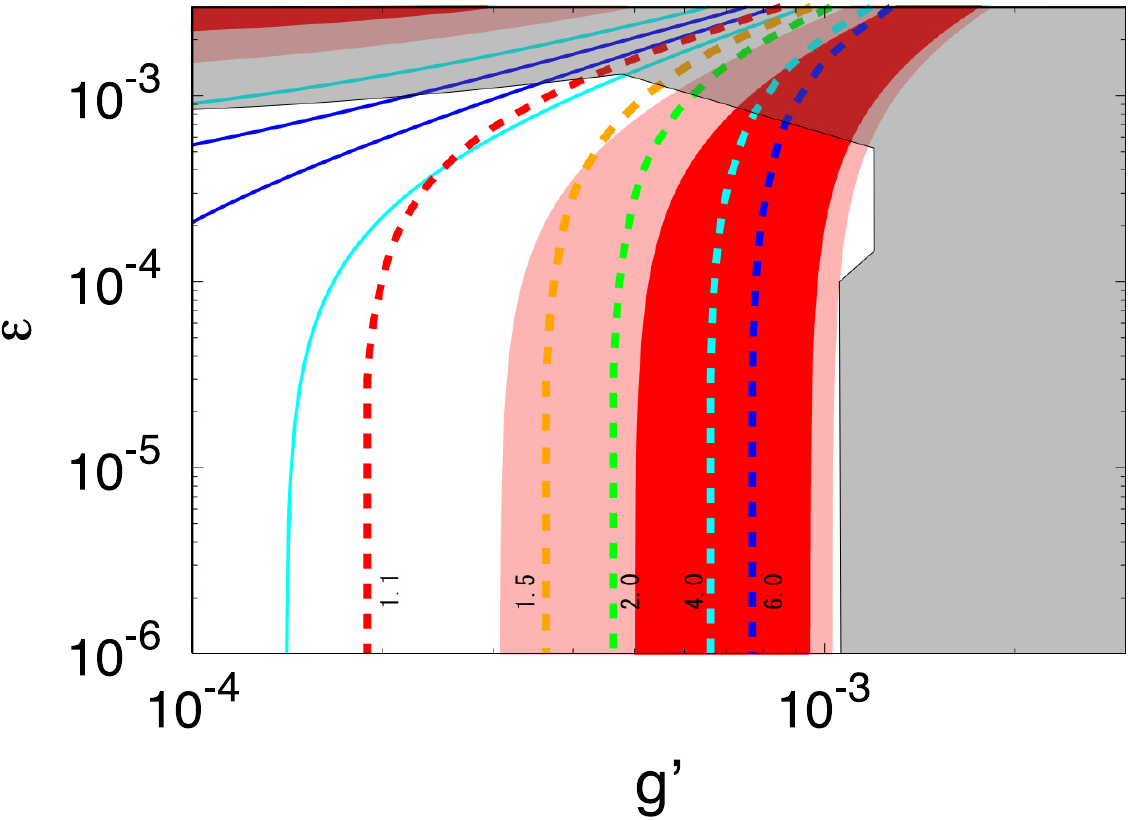}
&
\includegraphics[width=8.5cm]{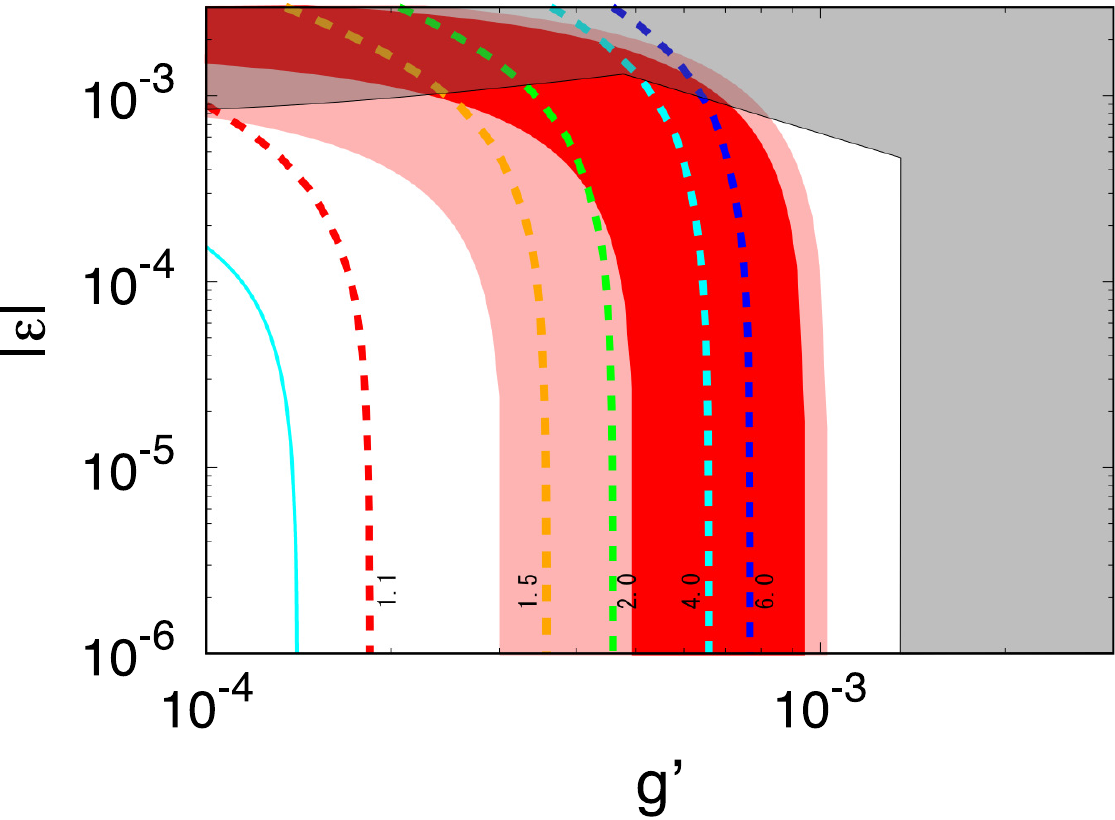} \\
\end{tabular}
\end{center}
\caption{ 
The contour plot of $R$ in the $g'$-$\epsilon$ plane for $10$ (top) and $50$ (bottom) MeV. The dashed curves 
represent $R$ with the number indicated aside. The grey region, the red and pink bands and the solid curves 
are the same in Figure \ref{fig:belle-g-2-1} and \ref{fig:belle-g-2-2}.
}
\label{fig:trident-1}
\end{figure}

\begin{figure}[t]
\begin{center}
\begin{tabular}{cc}
\includegraphics[width=8.5cm]{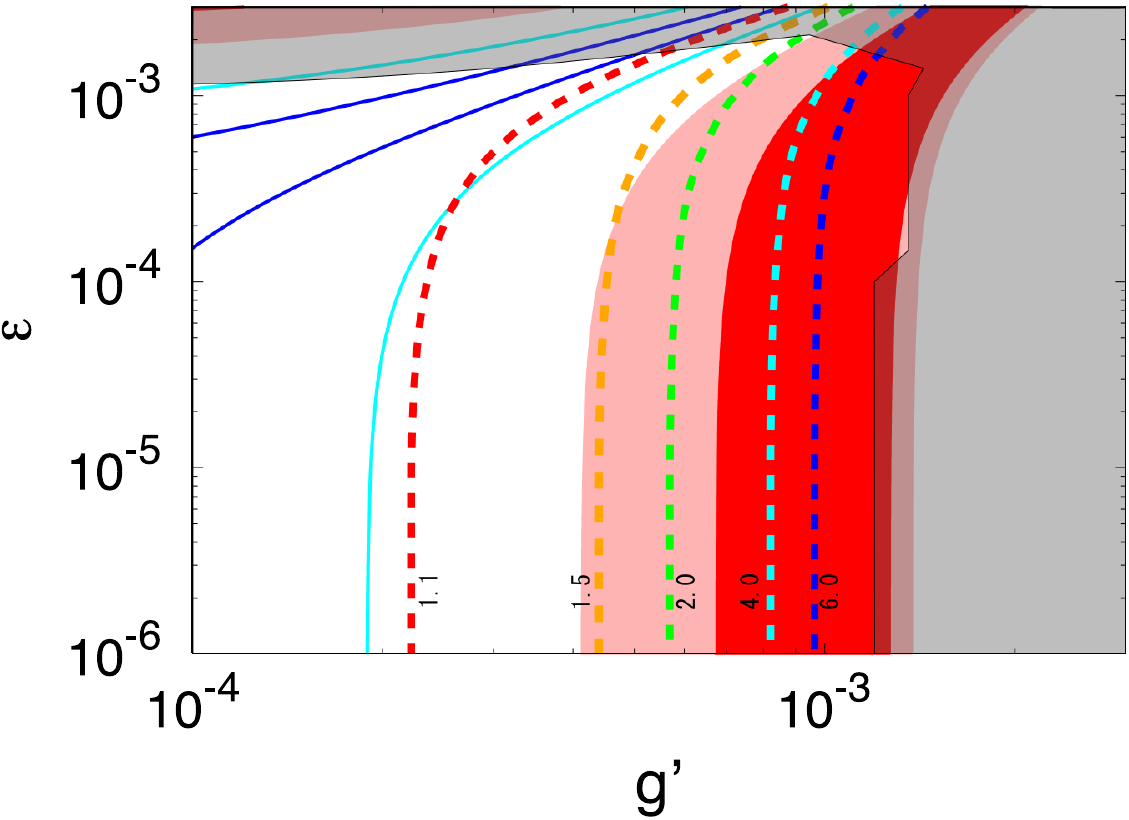}
&
\includegraphics[width=8.5cm]{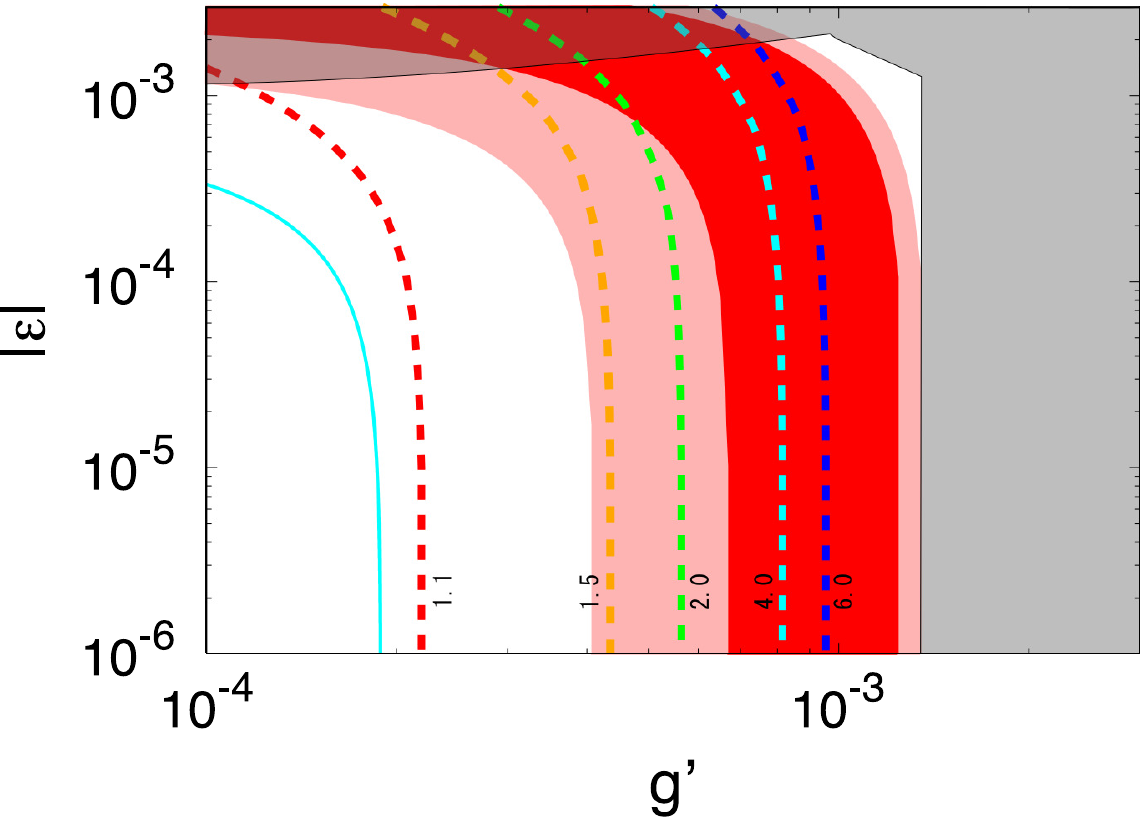} \vspace{2mm} \\
\includegraphics[width=8.5cm]{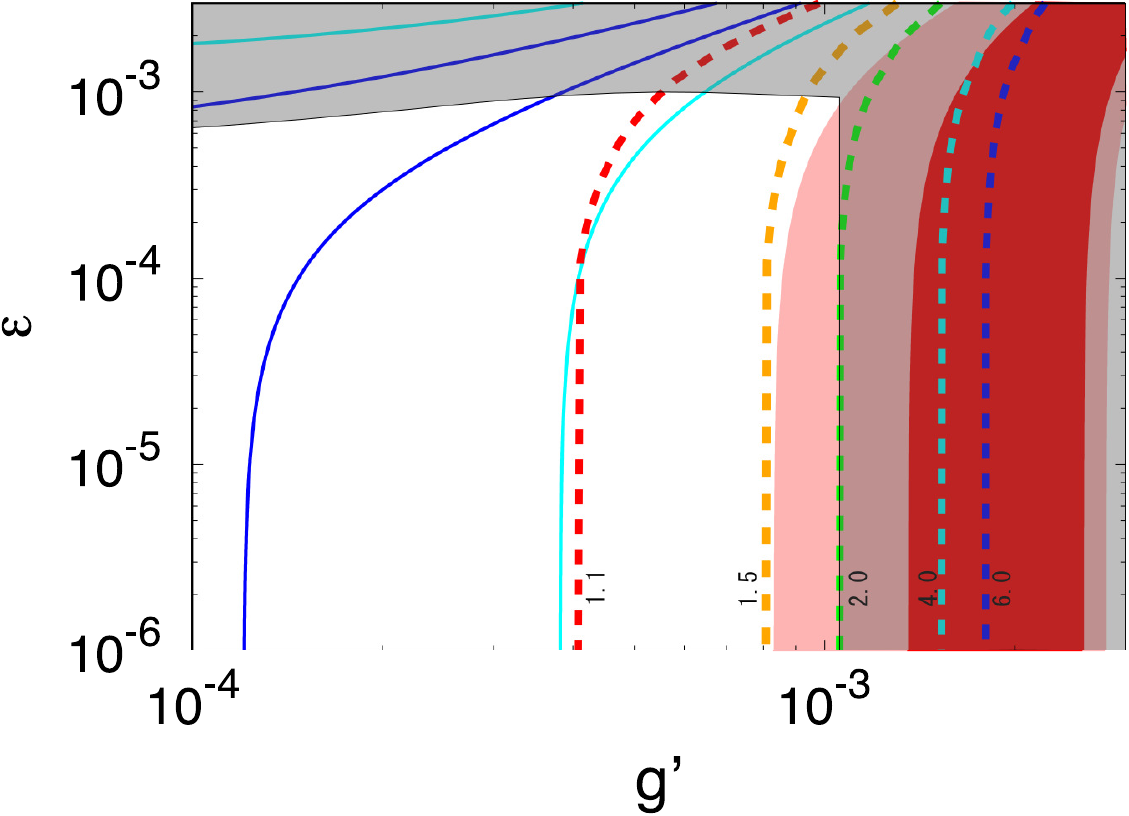}
&
\includegraphics[width=8.5cm]{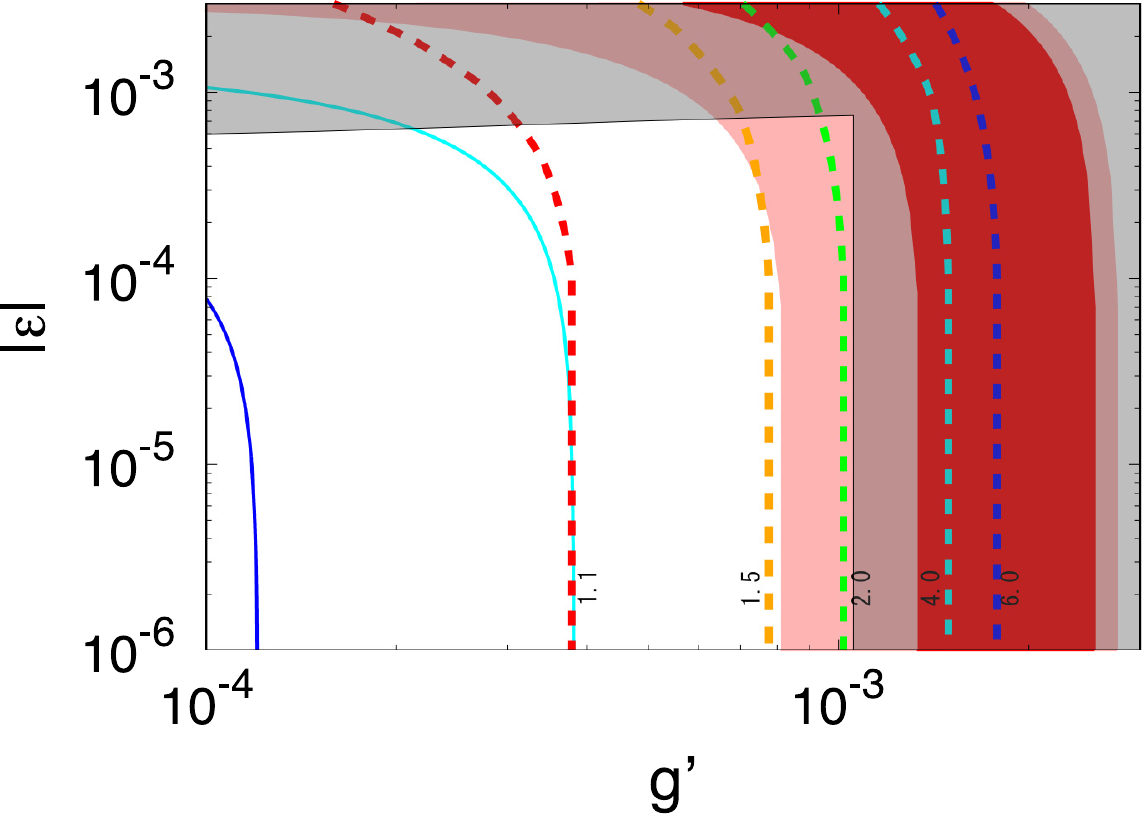} \\
\end{tabular}
\end{center}
\caption{ 
The same plots with Figure \ref{fig:trident-2} for $100$ (top) and $300$ (bottom) MeV.
}
\label{fig:trident-2}
\end{figure}

Figure \ref{fig:trident-CS} shows the cross section of the neutrino trident production (left) 
in the $L_\mu - L_\tau$ model and the SM, and the ratio of the cross  section to the SM one, $R$, (right) 
in terms of the neutrino energy, $E_\nu$.  We assume an iron target with the mass number $55.0$ and the atomic number $26$.
The kinetic mixing parameter is fixed to $\epsilon = 10^{-5}$, 
and the mass is chosen as $m_{Z'}=10$ (red curves) and $100$ (blue curves) MeV as reference values, respectively. 
The gauge coupling constant is taken to be  $g'=5.8 \times 10^{-4}$ (red-solid), $3.4 \times 10^{-4}$ (red-dashed), 
and $g'=9.5 \times 10^{-4}$ (blue-solid), $5.8 \times 10^{-4}$ (blue-dashed), respectively.
The grey curve represents the SM cross section. 
It is seen from the left panel that the trident production cross section becomes larger as the neutrino energy is larger.
It reaches to $(3.7$-$4.9)\times 10^{-40}$ cm$^2$ for $E_\nu = 100$ GeV while it does to $(0.12$-$1.0) \times 10^{-43}$ cm$^2$ 
at $E_\nu = 1$ GeV.
It is also seen from the right panel that the ratio $R$  
becomes larger as $E_\gamma$ is lower. This fact suggests that neutrino beams with lower energy have better sensitivity 
to search for the light $Z'$ boson. The ratio is roughly larger than $2$ for $E_\nu \lsim 1.5$ GeV for our reference parameters. 
Since the cross section becomes smaller for the lower energy beam, larger flux is inevitably needed to have enough events. 
For higher neutrino energies, such as DUNE \cite{Acciarri:2015uup} and SHiP \cite{ Anelli:2015pba}, 
the detailed study can be found in \cite{Magill:2016hgc}.

In figures \ref{fig:trident-1} and \ref{fig:trident-2}, the ratios of the cross section are shown for the same parameters of 
figures \ref{fig:allowed-region1} and \ref{fig:allowed-region2}. The values of $R$ are indicated near each curves. The 
energy of neutrino is taken to be $1.5$ GeV which is the same energy at the INGRID detector at the T2K experiment \cite{Abe:2011ks}. 
One can see that the contour curves are different in the left and right panels for each $m_{Z'}$. 
As explained in Sec. \ref{sec:allowed-region}, the difference originates from the relative phase between the amplitudes, 
and is significant for the lower neutrino energy. 
In the panels, it is seen that the region with $g'$ smaller than from the present bound can be searched even for $R \lsim 6$ 
except for $m_{Z'} = 300$ MeV. 
It is also seen that the same ratio as the CCFR experiment, $R\lsim1.1$, can provide the search for entire region of 
$(g-2)_\mu$ within $3\sigma$ for $m_{Z'} \lsim 300$ MeV and also some part of $\Delta a_\mu = 10^{-10}$.

\begin{figure}[t]
\begin{center}
\includegraphics[width=10cm]{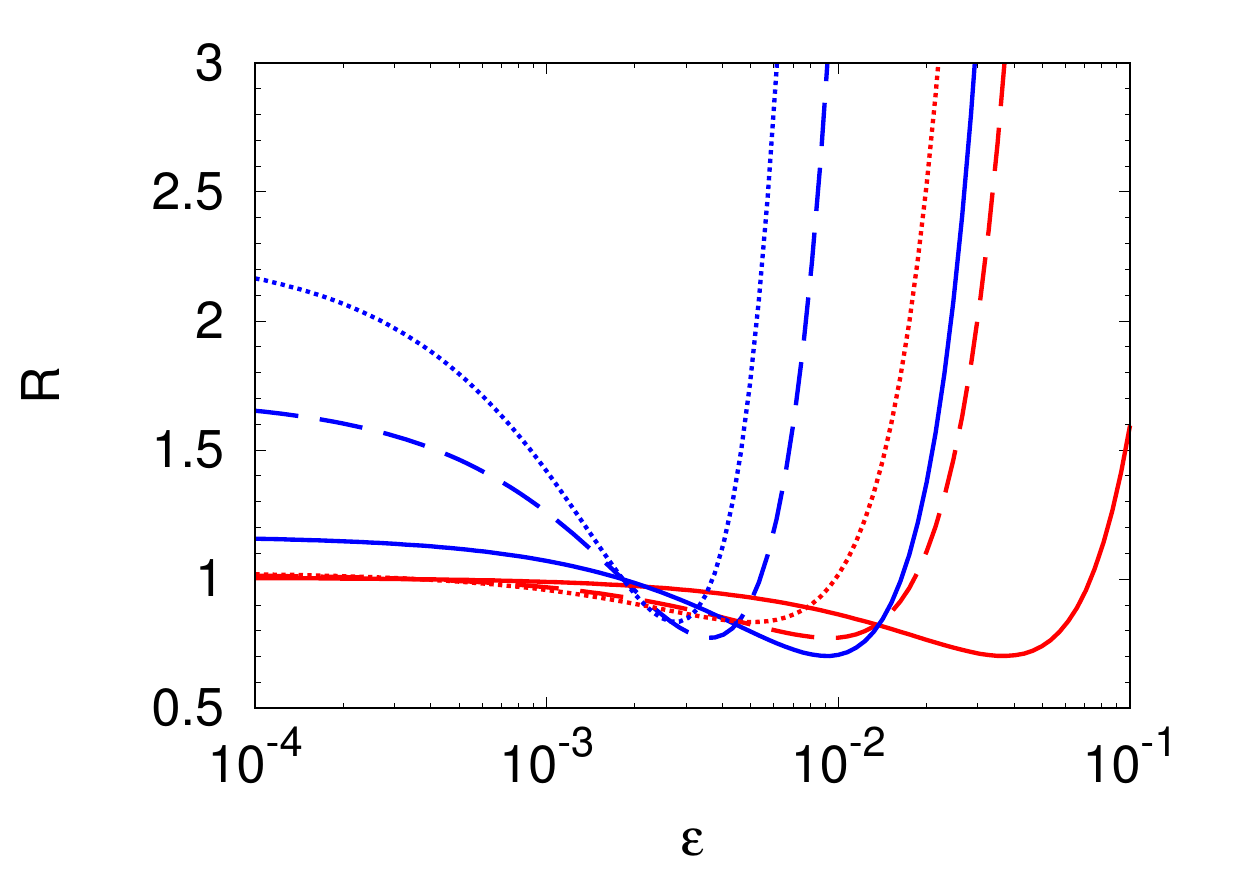}
\end{center}
\caption{ 
The ratio of the cross sections, $R$, as a function of $g'$. The solid, dashed and dotted curves represent $R$ for 
$m_{Z'} = 300,~100$ and $50$ MeV, and the red and blue ones for $g'=10^{-4}$ and $5 \times 10^{-4}$, 
respectively. The neutrino energy is fixed to $E_\nu = 1.5$ GeV. 
}
\label{fig:trident-3}
\end{figure}

As mentioned above and in Sec.~\ref{sec:allowed-region}, the $Z'$ contribution to the trident production 
cross section can be positive or negative depending on $\epsilon$. In fact, when $\epsilon > g' > 0$, the $Z'$ amplitude is negative and 
interferes destructively with the SM amplitude. 
Then, the ratio $R$ can become smaller than the unity. This can not happen 
in the $L_\mu - L_\tau$ model without the tree-level kinetic mixing because the loop induced kinetic mixing is always 
smaller than $g'$.

Figure \ref{fig:trident-3} shows the $\epsilon$ dependence of $R$. The mass of $Z'$ is taken to 
be $m_{Z'}=50,~100$ and $300$ MeV for the dotted, dashed and solid curves, and the coupling constant is taken to 
be $g'=10^{-4}$ and $5 \times 10^{-4}$ for the red and blue ones, respectively. 
The neutrino energy is fixed to $1.5$ GeV. 
It is seen that $R$ gradually decreases as $\epsilon$ increases, and then quickly increases after it reaches at a minimum. This behavior can be 
understood as follows. The interference term with the SM amplitude is proportional to $g'-\epsilon e \cos\theta_W$ 
while the absolute square of the $Z'$ amplitude is proportional to the square of that.  Therefore the total cross 
section decreases linearly in $\epsilon$. After reached at the minimum, the absolute squared term 
dominates over the interference term and the cross section increases quadratically in $\epsilon$. 
It is also seen that $\epsilon$ at the minimum is larger as $g'$ is smaller. Moreover, it can be seen that the 
minimum of $R$ is smaller for larger $m_{Z'}$ and is independent of $g'$. 
The minimum is easily obtained by minimizing the total cross section with respect to it, 
and is given by $ \epsilon_{\mathrm{min}}= \frac{1}{e\cos\theta_W}(g'+ g'^{-1} \frac{A}{B})$ where $A$ and $B$ are independent of $g'$ and $\epsilon$, and 
determined by the $Z'$ amplitude. 
Then, using $\epsilon_{\mathrm{min}}$, the minimum of $R$ is given by $R_{\mathrm{min}} = 1 - \frac{A^2}{\sigma_{\mathrm{SM}} B}$ 
where $\sigma_{\mathrm{SM}}$ stands for the SM cross section, which is independent of $g'$ as well as $\epsilon$. Therefore, the 
minimum is determined only by $m_{Z'}$ and $E_\nu$.

The neutrino trident production process is sensitive to the sign of $\epsilon$ for $|\epsilon| \gg g'$ while one photon 
plus missing search is insensitive to it. 
Thus, the neutrino beam experiment can provide the different information from the Belle-II experiment. 
For $|\epsilon| \ll g'$, the constraint becomes independent  of $\epsilon$, and hence tight bounds can be set on it.
On the other hand, the production cross section of $Z'$ at $e^+ e^-$ colliders is proportional to $\epsilon^2$ and 
hence it can not explore the small kinetic mixing region. Thus, the searches for the neutrino trident production process 
are complementary to the $e^+ e^-$ collider search, and are important to the search for the light and weakly interacting gauge 
boson.

\section{Summary and Discussion} \label{sec:summary}
We have considered the light and weakly interacting $Z'$ boson in the gauged $L_\mu - L_\tau$ model, 
simultaneously taking into account the gauge interaction and the kinetic mixing. 
We studied the possibilities on the search for such the $Z'$ boson analyzing one photon plus 
missing (neutrinos) events  at the Belle-II experiment 
and the neutrino trident production process at neutrino beam experiments. 

We have shown the allowed region in the $g'$-$\epsilon$ plane for $m_{Z'} = 10,~50,~100$ and $300$ MeV applying 
various experimental constraints and requirements. Then, the one photon plus missing events from $Z'$ decay were analyzed in the 
allowed region. We showed that the differential cross section in terms of $E_\gamma$ has an characteristic shape, and 
found the signal can be larger than the SM background for $|\epsilon| > 6.0 \times 10^{-6}$ at least at the edge of $E_\gamma$. 
Thus, the search for the light $Z'$ boson will be possible at the Belle-II experiment.
We also showed the cross section for the parameter space in the $g'$-$\epsilon$ plane that can be explored at the Bell-II experiment.

For the neutrino trident production process, we showed that a neutrino beam with lower energy is more sensitive to the existence of  $Z'$. 
Then, taking $E_\gamma = 1.5$ GeV, the sensitivity was shown in the $g'$-$\epsilon$ plane. 
We found that even the ratio $R \simeq 6$,  smaller parameters than the present bound can be explored. 
When the trident production cross section is measured more precisely, the whole region of 
$(g-2)_\mu$ favored region can be covered. We have shown that the neutrino trident production process is also sensitive 
to the sign of $\epsilon$ while the one photon plus missing search is not. Therefore both experiments will be complementary 
in searching for the light $Z'$ boson.

Before closing the summary, two comments are in order. 1) For the search at the Belle-II, 
$e^+ + e^- \rightarrow$ multi-$\gamma$ can also be serious backgrounds if several photons are undetected. 
The total cross sections of $2$-, $3$- and $4$-gamma final states are roughly estimated as $10^9,~10^8$ and $10^6$ 
ab, respectively. Thus, the expected numbers of the these backgrounds would be much larger than that of the signal 
events.  
For two photon in the final state, 
changing the cut on the photon angle will reduce this background. However, for more photons case, it is not easy to 
reduce such the events, especially for the cases that only one photon is measured and other photons escape to beam directions. Therefore more detailed study on the background is needed to determined the 
parameter space to be explored. 
2) For the neutrino trident production process, the momenta and angle distributions 
of the muons are important to discriminate the signal from the background. We leave these for our future works.

\begin{acknowledgements}
The authors would like to thank K.~Hayasaka and T.~Yoshinobu for fruitful discussions and useful information on the 
Belle-II detector. Y.K would like to thank the visitor support program in Japan Particle and Nuclear Forum.
The work of T.S is supported by JSPS KAKENHI Grants No.~15K17654. 
\end{acknowledgements}

\bibliographystyle{apsrev}
\bibliography{biblio}

\end{document}